\documentclass[12pt]{article}
\usepackage[tmargin=2cm,lmargin=2cm,rmargin=2cm,bmargin=2cm]{geometry}
\usepackage{setspace}
\usepackage{parskip}
\usepackage{amssymb}
\usepackage{amsmath}
\usepackage{graphicx}
\usepackage[english]{babel}
\usepackage{fancyhdr}
\usepackage[T1]{fontenc}
\usepackage{gensymb}
\usepackage{siunitx}
\usepackage{graphicx} 
\usepackage{float}
\usepackage{wasysym}
\usepackage{mathtools}
\usepackage{authblk}
\usepackage{titlesec}
\usepackage[export]{adjustbox}
\usepackage{textgreek}

\usepackage[
backend=bibtex,
hyperref=true,
bibencoding=utf8,
natbib=true,
maxbibnames=99,
style=nature,
citestyle=nature
]{biblatex}

\AtEveryBibitem{\clearfield{month}}
\AtEveryBibitem{\clearfield{issn}}
\AtEveryBibitem{\clearfield{isbn}}

\usepackage{hyperref}

\usepackage{amssymb}

\usepackage[dvipsnames]{xcolor}

\usepackage{amsmath}
\usepackage{listings}
\usepackage[T1]{fontenc}
\usepackage{geometry}

\usepackage{caption}
\usepackage{subcaption}

\usepackage[cp1252]{inputenc}

\usepackage{lmodern}
\usepackage{multirow}
\usepackage[flushleft]{threeparttable}

\newcommand{\modulename}[1]{\textit{#1}}

\newcommand{\SymPhas}{\textit{SymPhas}}
\newcommand{\CC}{C++}

\usepackage{inconsolata}

\begin{document}

\title{\SymPhas{}---General purpose software for phase-field, phase-field crystal and reaction-diffusion simulations}

\author[a,b]{Steven A. Silber}
\author[a,b,c*]{Mikko Karttunen}

\affil[a]{Department of Physics and Astronomy, The University of Western Ontario,  1151 Richmond Street, London, Ontario, Canada N6A\,3K7}
\affil[b]{Centre for Advanced Materials and Biomaterials Research, The University of Western Ontario,  1151 Richmond Street, London, Ontario, Canada N6A\,5B7}
\affil[c]{Department of Chemistry, The University of Western Ontario,  1151 Richmond Street, London, Ontario, Canada N6A\,5B7}


\affil[*]{Email: mkarttu@uwo.ca}

\maketitle

\begin{abstract}

This work develops a new open source API and software package called \SymPhas{} for simulations of phase-field, phase-field crystal and reaction-diffusion models, supporting up to three dimensions and an arbitrary number of fields. \SymPhas{} delivers two novel program capabilities: 1) User specification of models from the associated dynamical equations in an unconstrained form
and 2) extensive support for integrating user-developed discrete-grid-based numerical solvers into the API. The capability to specify general phase-field models is primarily achieved by developing a novel symbolic algebra functionality that can formulate mathematical expressions at compile time, is able to apply rules of symbolic algebra such as distribution, factoring and automatic simplification, and support user-driven expression tree manipulation.
A modular design based on the \CC{} template meta-programming paradigm is applied to the symbolic algebra library and general API implementation to minimize application runtime and increase the accessibility of the API for third party development. 
\SymPhas{} is written in C/\CC{} and emphasizes high-performance capabilities via parallelization with OpenMP and the \CC{} standard library. 
\SymPhas{} is equipped with a forward Euler solver and a semi-implicit Fourier spectral solver. 
Sample implementations and simulations of several phase-field models are presented, generated using the semi-implicit Fourier spectral solver. 
\end{abstract}

\section{Introduction}

The phase-field modeling technique involves finding and minimizing the free energy of one or more order parameters that describe a phase \cite{Hohenberg1977,Provatas2010}. The method was originally introduced by Fix in 1983 \cite{Fix1983} and has since been applied to modeling a wide range of problems including material microstructures \cite{Chen2002,Li2017}, crack propagation \cite{Spatschek_2011}, batteries \cite{Wang2020}, biological membranes \cite{Fan_2008}, cellular systems \cite{Nonomura_2012,Palmieri2015} and even immune response \cite{Najem2014}. Some recent advancements 
include the phase-field crystal model \cite{Elder2002} and its applications, see e.g., Refs.~\cite{Achim_2006,Emmerich_2011,Faghihi2013,Kocher_2019,Alster2020}, and phase-field damage models \cite{Wu_2017}. 

The same field-based approach can also used for problems in which a free energy description is not readily available. In such a case, the equations of motion are, instead, derived phenomenologically. A well-known class of such problems are reaction-diffusion systems, including Turing~\cite{Turing1952} and Gray--Scott models~\cite{Gray_1985}. These exhibit complex morphologies which mimic nature~\cite{Lee_1994,Leppaenen2002,Maini_2012} with far-reaching applications to biological systems~\cite{Murray1989}.

With the phase-field approach gaining popularity, there is a growing need for 
open source phase-field simulation software that generalizes some numerical strategies as has been discussed by Hong and Viswanathan~\cite{Hong2020}. Phase-field simulation software can generally be divided between numerical solvers using finite element or finite difference methods.
On the finite element side, some recent ones include, e.g.,
PRISMS-PF \cite{DeWitt2020}, which uses a matrix free approach, as well as
SfePy \cite{Cimrman_2019}, FEniCS \cite{Alnaes2015} and
MOOSE \cite{Permann2020}, of which the latter offers symbolic algebra functionality and automatic differentiation, allowing for instance, specification of free energy equations. The package FiPy \cite{Guyer2009} is equipped with an accessible Python interface and equation specification.

While there are several packages for the finite element method to solve partial differential equations,
much less is available for finite differences. 
Current open software includes
the Mesoscale Microstructure Simulation Project (MMSP) \cite{Keller2019}, though development appears to have ceased shortly after its release, and 
OpenPhase \cite{Tegeler2017}, which employs parallelization and sparse storage 
to solve large-scale multi-phase problems.

To improve upon the existing open source tools available using finite difference methods, we develop \SymPhas{}, an API and software package 
that aims at advancing the ability to implement numerical 
solutions of general phase-field problems by maximizing accessibility, flexibility and performance. \SymPhas{} uses a discrete-grid-based approach for the implementation of numerical solvers to allow
simulations to scale well with the number of grid points and number of order parameters, regardless of dimension. This is supplemented with parallelization via the \CC{} standard library and OpenMP~\cite{Dagum_1998}.

The \SymPhas{} API allows the user to define and solve any phase-field model that can be formulated field-theoretically, up to three dimensions and with arbitrary numbers of order parameters and equations. This extends to reaction-diffusion problems as well.
Phase-field problems are readily specified in the program using the dynamical equations provided in a completely unconstrained form using simple \CC{}-macro-based grammar. We achieve this primarily in two ways: 1) Development of a symbolic algebra library to manipulate and transform mathematical constructs as expression trees and 2) a modular approach of Object Oriented Programming (OOP) that progressively layers more complexity as needed by a given application.

The symbolic algebra feature is implemented as compile-time constructs that directly formulate expression trees at the point of definition. This is a unique feature of \SymPhas{} that is, to the best of our knowledge, not present in any other phase-field software package. 

A modular design is used to retain a simple interface for basic uses while simultaneously supporting complex tasks and implementations. This design applies template meta-programming to fully optimize program constructs and eliminate branching wherever possible. We also achieve considerable decoupling between elements, allowing individual functional elements of \SymPhas{} to remain distinct; this has the added benefit of supporting community development. 

The modular approach also facilitates another key feature of \SymPhas{}: The ability to integrate a user-developed numerical solver into the workflow. The solver is seamlessly integrated via a class inheritance strategy
designed to eliminate almost all API-specific restrictions on the solver implementation. In writing the solver, the user can leverage the significant set of features available in the symbolic algebra library and in the \SymPhas{} API overall.

Through extensive documentation and adherence to best programming practices, we provide \SymPhas{} as a codebase to be expanded and driven by community development.
This is further facilitated by managing the build process with CMake~\cite{cmake}, which provides \SymPhas{} with multi-platform support and grants users the ability to customize the compilation and installation procedure in a straightforward way.

\section{Methods}

To generate solutions to phase-field problems, an implementation that defines the problem and establishes the program control flow is written in \CC{} using the \SymPhas{} API. This consists of three components:

\begin{enumerate}
	\item \textbf{Model definitions file:} The phase-field description with the equations of motion are specified. These are written using \CC{} macros provided by \SymPhas{} and follow a simple grammar structure. Putting this in a separate file is optional.
	\item \textbf{Solver file:} The implementation of a specific method which solves a phase-field problem using the equations of motion.
	\item \textbf{Driver file:} Specifies the workflow, data inputs and outputs.
\end{enumerate}

We use OOP 
and extensively apply the programming paradigm
known as \textit{template meta-programming}; the use of objects and functions defined with arbitrary parameters or data types~\cite{Meyers2005}. 
These abstractions are either implicitly (by the compiler) or explicitly (by the user) specialized for concrete types. The benefit of this approach is that the type specialization gives the compiler full information about the call stack, allowing it to make optimizations not possible in different approaches (e.g. virtual inheritance). The other advantage is added extendability through type dependent implementations. The drawback is that since each specialization is unique, the library and executable will take longer to compile and result in a larger size when many specializations are used. For example, compiling five phase-field models of one or two order parameters with one solver defined with all available finite difference stencils -- see Section~\ref{methods:objects:stencils} -- results in a total size of approximately 3\,MB.
As part of template meta-programming, we also apply the \textit{expression template technique}~\cite{Veldhuizen1995,Vandevoorde2003}, commonly referred to as the \textit{curiously recurring template pattern} (CRTP) \cite{Coplien1996}, mainly used in the implementation of the symbolic algebra functionality.
A non-exhaustive list of familiar libraries using expression templates includes Armadillo~\cite{Sanderson2016,Sanderson2018}, Blitz++~\cite{Veldhuizen2000}, Boost $\mu$BLAS~\cite{Schaeling2011}, Dlib~\cite{King2009}, Eigen~\cite{GaeelGuennebaud2010}, Stan Math Library~\cite{Carpenter2015} and xtensor~\cite{xtensor}.

The OOP approach applies a modular design to program structure; this means that we minimize coupling and maximize cohesion~\cite{Vanderfeesten_2008, Candela2016} of all classes in the API, as well as designing each element to support class inheritance or composition.
The former is primarily accomplished by following best programming principles such as applying the single-responsibility principle for objects \cite{Martin2002}. The latter implies that objects designed under this modular framework can be readily extended or modified without refactoring the existing code. Moreover, modularity is used to reflect the real world representation of a phase-field problem. The overall aim is to simplify and streamline the future development of \SymPhas{}.

Additionally, the build process is another aspect designed to be user friendly. Managed by CMake, the user has full control over program definitions and modules. The result of the build process is a shared library that can be linked in a g++ invocation or alternatively, imported into a separate user CMake \cite{cmake} project.

\subsection{Overview of Modules}

There are four required and two supplementary modules as part of \SymPhas{}. 
The necessary modules constituting the \SymPhas{} library are the basic functionality (\modulename{lib}), datatypes (\modulename{datatypes}), solution (\modulename{sol}) and symbolic algebra (\modulename{sym}) modules. The module \modulename{lib} is a dependency of all other modules, since it introduces components such as objects and types used throughout the program.
There are two additional libraries which complete the feature set of \SymPhas{}: the configuration (\modulename{conf}) and the input/output (\modulename{io}) modules.

The \modulename{datatypes} module depends only on \modulename{lib}. It implements objects used in the discrete grid representations, the most important of which is the \lstinline{Grid} class, a managed array for storing data of 1-, 2- and 3-dimensional problems. 
The symbolic algebra library (\modulename{sym}) provides the core functionality 
to interpret mathematical expressions. The implementation of \modulename{sym} contains all the mathematical objects and relations that are required to specify a phase-field equation of motion. It also specifies rules between these objects.
The solution module (\modulename{sol}) provides the structural and functional framework used to describe and solve a phase-field problem. This is accomplished by defining two objects: One being the programmatic representation of a general phase-field problem and the other which implements a specific set of interface functions for time evolving the phase-field data.

\subsection{Objects in \SymPhas{}} \label{methods:objects}

The following is a list and brief description of the relevant objects
in \SymPhas{}:

\begin{itemize}
	\item \lstinline{Grid}: Basic array type for storing phase-field data of arbitrary type.
	\item \lstinline{Boundary}: Logical element for defining the properties of a grid boundary.
	\item \lstinline{Stencil}: Object which defines the finite difference stencils used to approximate derivatives in a uniform grid.
	\item \lstinline{OpExpression}: Interface object representing the node in an expression tree, based on CRTP.
	\item \lstinline{Solver}: Interface that is specialized for implementing the
	solution procedure.
	\item \lstinline{Model}: Encapsulation of the problem representation, including the equations of motion. Primary interface for managing the solution of a phase-field model.
\end{itemize}

\subsubsection{Uniform Grid}

The data of a grid is initialized in computer memory as a one-dimensional array, and the desired system dimension is logically imposed according to row major order where the first dimension is always the horizontal ($x$-axis). This ensures fastest run time through memory localization. 

An extension of \lstinline{Grid} called \lstinline{BoundaryGrid} enables the use of finite difference stencils at points near the boundary and implements routines to update boundary grid elements (for example, to apply periodic boundaries). This is accomplished by managing a list of indices that correspond to boundary elements. The number of layers of the boundary is predefined based on the extent of the largest finite difference stencil.

The design of the \lstinline{Grid} class and any specialization thereof is based on template meta-programming, and allows the user to select the data type and the dimension.

\subsubsection{Finite Difference Stencils} \label{methods:objects:stencils}

Stencils are finite difference approximations of derivatives of a specified order. In \SymPhas{}, we implement second and fourth order accurate central-space stencils for various orders of derivatives. To this end, stencils are defined using three characteristics: 1) The order of derivative that is approximated, 2) the order of accuracy,
and 3) the dimension of the system. An additional characterization is the number of points used in the approximation. A stencil family is a group of stencils with the same dimension and order of accuracy. We apply this categorization to the design of stencils in \SymPhas{} by implementing specialized template classes for each family.

Stencils are CRTP-based template classes with member functions for each order of derivative up to fourth order, and a member function for the generalized implementation of higher orders. In particular, the Laplacian, bilaplacian, gradlaplacian and gradient derivatives are explicitly defined according to those derived in \cite{Patra2006}, including both anisotropic- and isotropic-type stencils that are second and fourth order accurate in two dimensions, and second order accurate in three dimensions. Using CRTP in the stencil implementation eliminates branching in the member function invocations; a significant optimization that improves performance since stencils are applied multiple times at each grid point for every solution iteration. 

For second order accuracy approximations, the Laplacian is implemented by 5 and 9 point stencils for 2D, and 7, 15, 19, 21 and 27 point stencils in 3D; the gradlaplacian is implemented by 6, 8, 12 and 16 point stencils, and 10, 12, 28, 36 and 40 point stencils in 3D; and the bilaplacian is implemented by 13, 17 and 21 point stencils for 2D, and 21, 25, 41, 52 and 57 point stencils for 3D. For fourth order accuracy approximations, which are only in 2D, the Laplacian is implemented by 9, 17 and 21 point stencils; the gradlaplacian is implemented by 14, 18, 26 and 30 point stencils. The bilaplacian is implemented by 21, 25, 33 and 37 point stencils. The wide selection ensures that appropriate approximations can be used a problem as necessary.

\subsubsection{Expressions} \label{methods:objects:expression}

One of the primary features of \SymPhas{} is the symbolic algebra library, which represents mathematical expressions with expression trees formulated at compile time. An expression tree representation allows the equations of motion to be treated as a single object, i.e., they can be persisted as an object state and passed as a function parameter. Moreover, expression trees provide the ability to reorganize and manipulate mathematical expressions. 
The symbolic algebra functionality is used to interpret equations of motion that are provided in a general form, and is exposed in a user-friendly way via the \SymPhas{} API. 

Symbolic algebra is implemented in \SymPhas{} through an approach unique to phase-field simulations programs: CRTP is applied to generate compiled code for an expression tree evaluation so that it is as close as possible to writing the evaluation manually. In this regard, the symbolic algebra is considered ``compile-time constant'', since the expression representation is formulated at compile time. The type name of the CRTP base and expression tree node is \lstinline{OpExpression}.

One motivation of this design choice is minimizing application runtime, since a design that applies a deterministic control flow to expression tree traversal can significantly increase performance. 
The most significant corresponding improvement in performance is the reduction in the time spent by a numerical solver in evaluating the equation of motion of the phase-field model, which takes place for all points in the grid and for each iteration of the solver. 
An implication
is that, in general, each expression is a unique type (unique CRTP specializations of \lstinline{OpExpression}).

\subsubsection{Solution Interface}

The base \SymPhas{} library consisting of the six aforementioned modules does not contain a solver implementation, though two solver implementations that are detailed in Section~\ref{methods:implementation} are provided in the \SymPhas{} package obtained from Github (\lstinline{https://github.com/SoftSimu/SymPhas}). Instead, the \textit{solution interface}, \lstinline{Solver}, declares three functions 
which a concrete numerical solver must implement.
The solver interface design uses CRTP and is based on applying a mediator design pattern via a special object that we refer to as the ``equation mediator object''. These are generated by the solver for each dynamical equation in a phase-field model before the simulation begins. Its purpose is to recast the equation of motion into a form that can be interpreted by the numerical scheme of the solver in order for the subsequent time index of the corresponding phase-field data to be computed. The equation mediator object is constructed by the solver member function \lstinline{form_expr_one()}.
By taking advantage of the modular framework,
we design the solver interface to allow the equation mediator object to remain entirely specific to the implemented solver, maximizing third party development potential.
A specialized solver implements the following three primary interface functions, where additional functions, such as derivatives, may also be written as necessitated:
\begin{itemize}
	\item \lstinline{form_expr_one()}: Given the set of equations of motion of a phase-field model, constructs the equation mediator object for a specified equation of motion.
	This function is only called once. It performs as much computational work as possible to ensure maximum program performance.
	\item \lstinline{equation()}: Using the phase-field data and the equation mediator objects, performs an initial time evolution step,
	typically writing intermediate results to working memory.
	\item \lstinline{step()}: Using the phase-field data and intermediate results computed by \lstinline{equation()}, obtain the next iteration in the solution.
\end{itemize}

\subsubsection{Problem Encapsulation}

To represent the physical phase-field problem in the code domain, objects of basic functionality are successively encapsulated to build necessary functionality. For instance, the \lstinline{Grid} class is encapsulated by \lstinline{System} to add information such as the spatial intervals and discretization width. A further encapsulation will specialize \lstinline{System} into \lstinline{PhaseFieldSystem}, adding functionality such as data persistence and the ability to populate array values with initial conditions.
Using template meta-programming, the\\ \lstinline{PhaseFieldSystem} object allows the user to select the data type and dimension used for the instantiated type. It also
allows encapsulating \lstinline{Grid} or any of its specializations to modify the basic implementation features as required by the problem or solver. 

All of the aspects which constitute a phase-field model are encapsulated in the class \lstinline{Model}, of which the responsibility is to manage the phase-field data and numerical solver and initialize all phase-field data in a standardized way. It is also the primary interface to the phase-field data and interacting with the solver. 
\lstinline{Model} is itself specialized in order to manage a specific set of equations of motion corresponding to a phase-field problem. The user is responsible for this final specialization through the procedure defined in Section~\ref{methods:capabilities:models}.

\subsection{Capabilities} \label{methods:capabilities}

To produce solutions to 
phase-field problems, \SymPhas{} offers a number of capabilities.
These include convenient parameter specification alongside a rich feature set for specifying the equations of motion and managing the phase-field problem. In this section, we briefly list the capabilities relevant in typical use cases and outline the steps for generating a simple driver file in \SymPhas{}.

\subsubsection{Symbolic Algebra} \label{methods:capabilities:algebra}

Data are used in the symbolic algebra by linking it with a specialized expression type that represents a ``variable'' term (e.g., in the context of an equation of motion, each order parameter is a ``variable'' linked to the respective phase-field data).
The symbolic algebra also defines value literals and spatial derivative operators. Value literals include special constructs representing the positive and negative multiplicative identity (the numbers $1$ and $-1$) and the additive identity (the number $0$). These are mainly used to facilitate symbolic algebra rules. Some common functions are defined as well, including $\sin$, $\cos$ and the exponential. For less common cases, the convolution operator is also defined. Since the structure of expression trees is managed at compile time, type-based rules defined by specific expression tree structures are applied to formulate expressions, such as when addition, subtraction, multiplication and division operations are used. Rules include simplification, distribution, factorization.

While the primary purpose of the symbolic algebra feature is to represent an equation of motion for a phase-field problem and support the user in implementing a specialized solver, the same functionality can be applied in more general applications. 
Features such as the ability to name variables and print formatted expressions to an output stream in either simple text or \LaTeX{} format are included. Moreover, the symbolic algebra can be used to perform high-performance pointwise operations on arrays, include new symbols to the algebra ruleset, and even define identities for the new and existing symbols that would be applied automatically.

\subsubsection{Defining New Phase-Field Models} \label{methods:capabilities:models}

\SymPhas{} provides convenient \CC{} macro-based grammar to allow a user to define a new phase-field model in a completely unconstrained way, a novel feature among phase-field simulations software. Each new definition generates a specialization of the \lstinline{Model} class. Upon recompiling the program, the new model is fully functional without the pitfalls of verbose implementation details. Moreover, the same definition can be interpreted by all solvers available in \SymPhas{}.

A model is defined in three parts:
1) The model name as it appears in the compilation unit, which must be unique for all models, 2) a list of the order parameters and their types, and 3) the equations of motion. An additional optional section between the order parameter type list and equations of motion can be specified to define virtual variables. These can be used to measure desired system quantities or used in the equations of motion to optimize the runtime by pre-computing values. An example of defining a two-phase model with a virtual variable is demonstrated in Figure~\ref{fig:model_definition}. Specific details including all available macros are offered in the manual.

\begin{figure}
    \centering
    \footnotesize
    \begin{lstlisting}
MODEL(MC, 
  (SCALAR, SCALAR),
  PROVISIONAL_DEF(
    (SCALAR),
    var(1) = c5 * op(1) * op(2))
  MODEL_PREAMBLE_DEF(
    ( auto op13 = c2 * op(1) * op(1) * op(1);
      auto op23 = c4 * op(2) * op(2) * op(2); )
    dop(1) = lap(op(1)) + c1 * op(1) - op13 + lit(2.) * var(1),
    dop(2) = -bilap(op(2)) - lap(c3 * op(2) - op23 + c5 * op(1) * op(1)))
)
    \end{lstlisting}
    \caption{An example of specifying a phase-field model. Model~C~\cite{Hohenberg1977}, which represents eutectic growth with two order parameters~\cite{Elder1994}, is implemented. It is associated with the given name ``\lstinline{MC}'', which defines the type alias of the model in the code and therefore must be unique. The keywords \lstinline{op(N)} and \lstinline{dop(N)} refer to the \lstinline{N}th order parameter and its time derivative, respectively, and the keyword \lstinline{var(N)} refers to the \lstinline{N}th virtual variable. The keyword \lstinline{SCALAR} specifies that the field types are real-valued. The keyword \lstinline{lit(v)} is used to represent a numeric constant of value \lstinline{v} in the symbolic algebra expression. The keywords \lstinline{lap} and \lstinline{bilap} apply the 2nd and 4th spatial derivative to their arguments, respectively. The enumerated terms \lstinline{c1} to \lstinline{c5} are parameters passed to the model upon instantiation. Variables can be defined before the equations of motion using the macro \lstinline{MODEL_PREAMBLE_DEF}, demonstrated here with the cubic terms, \lstinline{op13} and \lstinline{op23}. This option exists chiefly for convenience and does not affect the structure of the expression tree formulated for the equation of motion. If this section is omitted and only the dynamical equations are provided, then the macro \lstinline{MODEL_DEF} is used instead.}
    \label{fig:model_definition}
\end{figure}

\subsubsection{Creating Custom Solvers}

The user may develop their own solver using the \SymPhas{} API and seamlessly integrate it into an existing workflow. Implementation of a solver entails extending the provided \lstinline{Solver} interface. 
A major benefit of the design is provided by the equation mediator object, since it remains completely internal to the functionality of the implemented solver and does not interact with the other parts of the API or program. This allows the solver implementation to be decoupled as much as possible from the surrounding implementation and allows the user to leverage the capabilities of the API without being limited by extensive requisite knowledge. 
Additionally, if the built-in \lstinline{SolverSystem} and its specializations are insufficient,
the user may develop a new specialization. This has some constraints and requirements, including following a specific naming style, inheritance requirements and recompilation of the solution module \modulename{sol}.

\subsubsection{Standardized Problem Parameter Management}

The parameters of the problem are managed by a specialized object tasked with managing data that fully describes a phase-field problem. 
This information includes the initial conditions, the interval data and information about the boundary conditions, the latter which can be specified on an individual basis.
There are a number of phase-field initialization routines available to the user, each tuned by user-provided parameters. Initialization routines defined by the user can also be used. 

This approach allows
the user to have a unified approach to initializing, accessing, and passing problem information, simplifying the workflow and ensuring flexibility.

\subsubsection{Input/Output}

The configuration module provides the user with the ability to write a configuration file with phase-field problem and data persistence parameters, which can be used to construct the problem parameters object.

The \SymPhas{} API includes data persistence capabilities through the \modulename{io} module, which introduces functions and objects for reading and writing phase-field data. The user is also provided with the ability to persist phase-field data at regular checkpoints throughout the simulation, which can later be used to recover simulation data from the last saved point if it is interrupted for any reason. This supports program reliability and convenience, particularly for extended simulations. 
Currently, there are three output/input formats: 1) Plain text matrix (the matrix format is amenable to plotting utilities such as gnuplot), 2) plain text column (an ordered list of vectors and values), and 3) binary output in the xdrfile format, popularized by GROMACS \cite{Lindahl2021}. This functionality is available to the user when \modulename{io} is compiled with \SymPhas{} through CMake.

Input can also be given to \SymPhas{} through the command line; this method allows configuring some program-level parameters. Unlike the configuration file, this a base component of \SymPhas{}. Command line parameters allow \SymPhas{} to change some of its behavior, largely with regards to the initial condition generation. The basic set of program level parameters are introduced in \modulename{lib}, and \modulename{io} introduces more parameters. The list of all configurable parameters and their details are provided in the manual.

\begin{figure}
	\centering
	\scriptsize
	\begingroup%
	\makeatletter%
	\providecommand\color[2][]{%
		\errmessage{(Inkscape) Color is used for the text in Inkscape, but the package 'color.sty' is not loaded}%
		\renewcommand\color[2][]{}%
	}%
	\providecommand\transparent[1]{%
		\errmessage{(Inkscape) Transparency is used (non-zero) for the text in Inkscape, but the package 'transparent.sty' is not loaded}%
		\renewcommand\transparent[1]{}%
	}%
	\providecommand\rotatebox[2]{#2}%
	\newcommand*\fsize{\dimexpr\f@size pt\relax}%
	\newcommand*\lineheight[1]{\fontsize{\fsize}{#1\fsize}\selectfont}%
	\ifx\svgwidth\undefined%
	\setlength{\unitlength}{419.73992097bp}%
	\ifx\svgscale\undefined%
	\relax%
	\else%
	\setlength{\unitlength}{\unitlength * \real{\svgscale}}%
	\fi%
	\else%
	\setlength{\unitlength}{\svgwidth}%
	\fi%
	\global\let\svgwidth\undefined%
	\global\let\svgscale\undefined%
	\makeatother%
	\begin{picture}(1,0.55151926)%
		\lineheight{1}%
		\setlength\tabcolsep{0pt}%
		\put(0,0){\includegraphics[width=\unitlength,page=1]{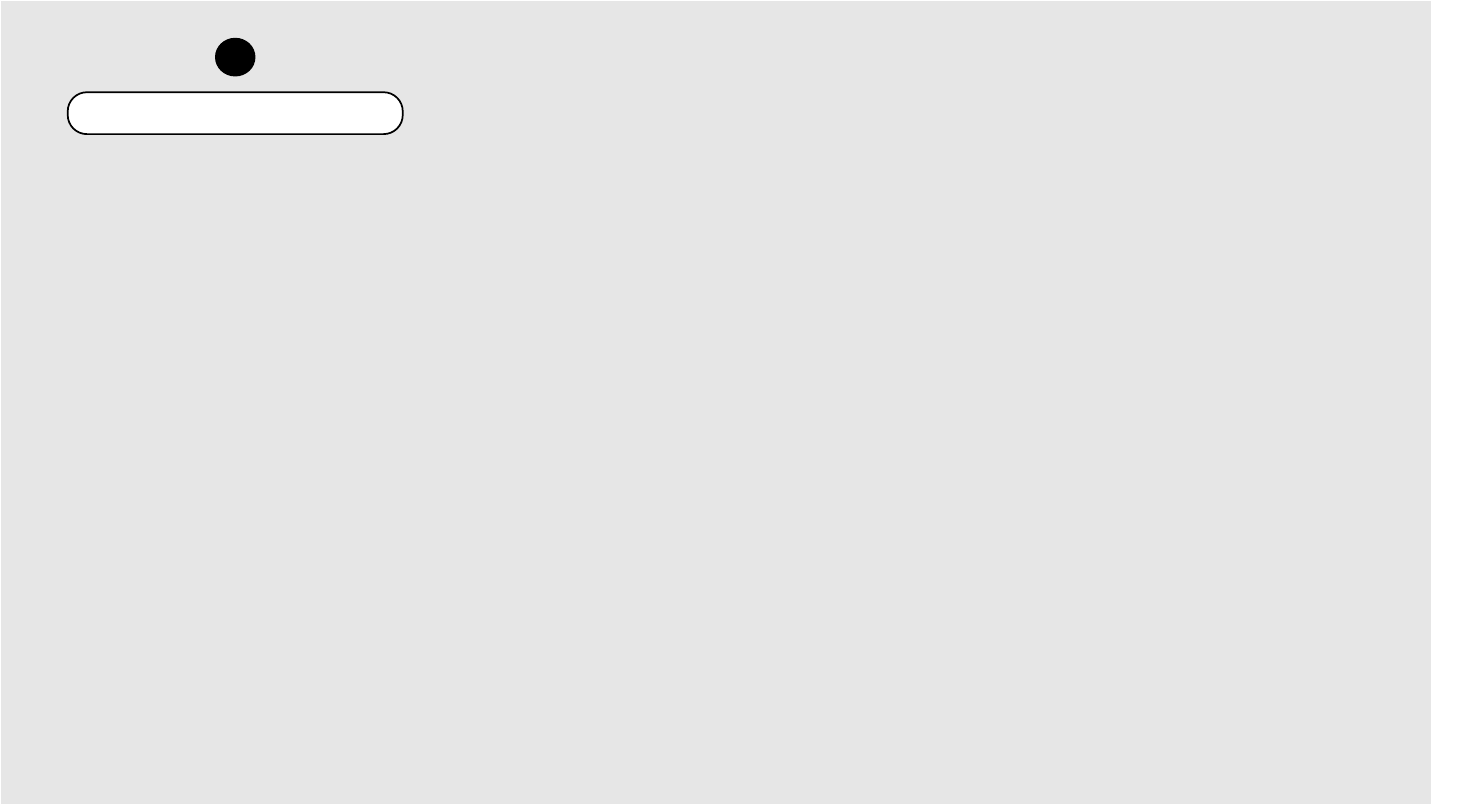}}%
		\put(0.16157335,0.46704034){\color[rgb]{0,0,0}\makebox(0,0)[t]{\lineheight{1.25}\smash{\begin{tabular}[t]{c}Initialize configuration\end{tabular}}}}%
		\put(0,0){\includegraphics[width=\unitlength,page=2]{implementation_small_v2.pdf}}%
		\put(0.49225814,0.5319947){\color[rgb]{0,0,0}\makebox(0,0)[t]{\lineheight{1.25}\smash{\begin{tabular}[t]{c}Driver\end{tabular}}}}%
		\put(0.82377576,0.5319947){\color[rgb]{0,0,0}\makebox(0,0)[t]{\lineheight{1.25}\smash{\begin{tabular}[t]{c}\lstinline{Model}\end{tabular}}}}%
		\put(0,0){\includegraphics[width=\unitlength,page=3]{implementation_small_v2.pdf}}%
		\put(0.16104047,0.42347701){\color[rgb]{0,0,0}\makebox(0,0)[t]{\lineheight{1.25}\smash{\begin{tabular}[t]{c}[Load Checkpoint?]\end{tabular}}}}%
		\put(0.48547551,0.10795121){\color[rgb]{0,0,0}\makebox(0,0)[lt]{\begin{minipage}{0.67051006\unitlength}\raggedright \end{minipage}}}%
		\put(0.16333496,0.5319947){\color[rgb]{0,0,0}\makebox(0,0)[t]{\lineheight{1.25}\smash{\begin{tabular}[t]{c}Solve Phase-Field Problem\end{tabular}}}}%
		\put(0,0){\includegraphics[width=\unitlength,page=4]{implementation_small_v2.pdf}}%
		\put(0.49065199,0.39721209){\color[rgb]{0,0,0}\makebox(0,0)[t]{\lineheight{1.25}\smash{\begin{tabular}[t]{c}Define problem parameters\end{tabular}}}}%
		\put(0,0){\includegraphics[width=\unitlength,page=5]{implementation_small_v2.pdf}}%
		\put(0.16099403,0.34815596){\color[rgb]{0,0,0}\makebox(0,0)[t]{\lineheight{1.25}\smash{\begin{tabular}[t]{c}Read backup file\end{tabular}}}}%
		\put(0,0){\includegraphics[width=\unitlength,page=6]{implementation_small_v2.pdf}}%
		\put(0.48967667,0.34821582){\color[rgb]{0,0,0}\makebox(0,0)[t]{\lineheight{1.25}\smash{\begin{tabular}[t]{c}Initialize \lstinline{Model} list\end{tabular}}}}%
		\put(0.49295187,0.22960868){\color[rgb]{0,0,0}\makebox(0,0)[t]{\lineheight{1.25}\smash{\begin{tabular}[t]{c}Select next \lstinline{Model}\end{tabular}}}}%
		\put(0.23460284,0.38343563){\color[rgb]{0,0,0}\makebox(0,0)[lt]{\lineheight{1.25}\smash{\begin{tabular}[t]{l}\textit{No}\end{tabular}}}}%
		\put(0.16829113,0.37616843){\color[rgb]{0,0,0}\makebox(0,0)[lt]{\lineheight{1.25}\smash{\begin{tabular}[t]{l}\textit{Yes}\end{tabular}}}}%
		\put(0.49056795,0.31364469){\color[rgb]{0,0,0}\makebox(0,0)[t]{\lineheight{1.25}\smash{\begin{tabular}[t]{c}[All models simulated?]\end{tabular}}}}%
		\put(0,0){\includegraphics[width=\unitlength,page=7]{implementation_small_v2.pdf}}%
		\put(0.49659382,0.26355095){\color[rgb]{0,0,0}\makebox(0,0)[lt]{\lineheight{1.25}\smash{\begin{tabular}[t]{l}\textit{No}\end{tabular}}}}%
		\put(0.43775105,0.27644197){\color[rgb]{0,0,0}\makebox(0,0)[lt]{\lineheight{1.25}\smash{\begin{tabular}[t]{l}\textit{Yes}\end{tabular}}}}%
		\put(0.82129282,0.23042956){\color[rgb]{0,0,0}\makebox(0,0)[t]{\lineheight{1.25}\smash{\begin{tabular}[t]{c}Update system (\lstinline{update})\end{tabular}}}}%
		\put(0,0){\includegraphics[width=\unitlength,page=8]{implementation_small_v2.pdf}}%
		\put(0.82129282,0.18497542){\color[rgb]{0,0,0}\makebox(0,0)[t]{\lineheight{1.25}\smash{\begin{tabular}[t]{c}Compute dynamics (\lstinline{equation})\end{tabular}}}}%
		\put(0,0){\includegraphics[width=\unitlength,page=9]{implementation_small_v2.pdf}}%
		\put(0.82129282,0.1395214){\color[rgb]{0,0,0}\makebox(0,0)[t]{\lineheight{1.25}\smash{\begin{tabular}[t]{c}Time evolve (\lstinline{step})\end{tabular}}}}%
		\put(0,0){\includegraphics[width=\unitlength,page=10]{implementation_small_v2.pdf}}%
		\put(0.82110556,0.09609849){\color[rgb]{0,0,0}\makebox(0,0)[t]{\lineheight{1.25}\smash{\begin{tabular}[t]{c}[Reached final index?]\end{tabular}}}}%
		\put(0,0){\includegraphics[width=\unitlength,page=11]{implementation_small_v2.pdf}}%
		\put(0.82148597,0.01546429){\color[rgb]{0,0,0}\makebox(0,0)[t]{\lineheight{1.25}\smash{\begin{tabular}[t]{c}Persist solution data\end{tabular}}}}%
		\put(0,0){\includegraphics[width=\unitlength,page=12]{implementation_small_v2.pdf}}%
		\put(0.78188778,0.34903378){\color[rgb]{0,0,0}\makebox(0,0)[t]{\lineheight{1.25}\smash{\begin{tabular}[t]{c}Model parameters\end{tabular}}}}%
		\put(0,0){\includegraphics[width=\unitlength,page=13]{implementation_small_v2.pdf}}%
		\put(0.76108252,0.05424335){\color[rgb]{0,0,0}\makebox(0,0)[lt]{\lineheight{1.25}\smash{\begin{tabular}[t]{l}\textit{No}\end{tabular}}}}%
		\put(0.83023066,0.04574204){\color[rgb]{0,0,0}\makebox(0,0)[lt]{\lineheight{1.25}\smash{\begin{tabular}[t]{l}\textit{Yes}\end{tabular}}}}%
		\put(0,0){\includegraphics[width=\unitlength,page=14]{implementation_small_v2.pdf}}%
	\end{picture}%
	\endgroup%

	\caption{Control flow diagram for the \SymPhas{} driver used to run simulations in this work. A list of models is generated using parameters from a configuration file. The number of models that is initialized depends on the configuration. 
	The \SymPhas{} library provides a function to perform the solution loop illustrated under \lstinline{Model}.
	}
	\label{fig:flow4}
\end{figure}%
\begin{figure}
    \centering
    \footnotesize
    \begin{lstlisting}
#include "symphas.h"
#define psi op(1)
#define dpsi dop(1)
MODEL(EX, (SCALAR), MODEL_DEF(
  dpsi = lap(psi) + (c1 - c2 * psi * psi) * psi))
  
int main(int argc, char* argv[]) {
  double dt = 0.5;
  symphas::problem_parameters_type pp{ 1 };
  symphas::b_data_type bdata;
  symphas::interval_data_type vdata;
  symphas::init_data_type tdata{ Inside::UNIFORM, { -1, 1 } };
  
  symphas::interval_element_type interval;
  interval.set_interval_count(0, 80, 128);
  
  bdata[Side::LEFT] = BoundaryType::PERIODIC;
  bdata[Side::RIGHT] = BoundaryType::PERIODIC;
  bdata[Side::TOP] = BoundaryType::PERIODIC;
  bdata[Side::BOTTOM] = BoundaryType::PERIODIC;
  vdata[Axis::X] = interval;
  vdata[Axis::Y] = interval;
  
  pp.set_boundary_data(&bdata);
  pp.set_initial_data(&tdata);
  pp.set_interval_data(&vdata);
  pp.set_problem_time_step(dt);

  model_EX_t<2, SolverSP<Stencil2d2h<5, 9, 6>>> model{ pp };
  symphas::find_solution(model, dt, 100);
}
    \end{lstlisting}
    \caption{Example of a simple driver program which includes the model definition.
    A phase-field problem of one order parameter named \lstinline{EX} is defined. System boundaries are defined to be periodic on each edge and initial conditions are set by seeding valuing with the uniform distribution $\mathcal{U}(-1, 1)$. Identical $x$ and $y$ intervals are defined and provided, resulting in a square $128\times128$ grid. When the model is created, the solver (\lstinline{SolverSP}) is passed as a template type parameter and the \lstinline{find_solution} function is called to perform 100 solver iterations. This program can be compiled through CMake or by providing the directories of the installed header and library to gcc.}
    \label{fig:driver_implementation}
\end{figure}

\subsection{Implementation} \label{methods:implementation}

The entire \SymPhas{} API is available on Github (\texttt{https://github.com/SoftSimu/SymPhas}) alongside two solvers (forward Euler, Sec.~\ref{sec:feuler} and the Semi-Implicit Fourier Spectral, Sec.~\ref{sec:spectral}), model definitions and driver file examples.
The program control flow of the driver file
used to to generate the simulations in this paper is illustrated in Figure~\ref{fig:flow}, and is found in the directory \lstinline{examples/simultaneous-configs} relative to the source code root. 
This driver performs several steps, including data persistence, as part of data collection, but a fully functional driver file can be as simple as is written in Figure~\ref{fig:driver_implementation} and included with the source code in \lstinline{examples/simple-driver}. It is presented here to illustrate relevant \SymPhas{} API elements, but primarily demonstrate its ease of use.

\subsubsection{Forward Euler Solver}
\label{sec:feuler}

A forward Euler solver is a well-known  explicit numerical method for partial differential equations. It is a first order method and it is known to have instabilities especially when solving stiff systems. It is provided as a base method. 
Since it is well-known and taught in virtually every course in numerical methods, we will not discuss it further but refer the reader to one of the standard references such as Press \textit{et al.}~\cite{Press1992}.

\subsubsection{Semi-Implicit Spectral Solver}
\label{sec:spectral}

Consider a phase-field problem for the order parameter $\psi = \psi(\vec{x},t)$; the equation of motion for this problem may be expressed in the form
\begin{equation}
	\frac{\partial \psi}{\partial t} = \mathcal{L}(\nabla^{n})\left\{\psi\right\} + \sum_i{{\mathcal{N}}_i(\nabla^{m_i})\left\{ f_i(\psi)\right\}}\,,
	\label{eq:spectralphasefield}
\end{equation}
where $\mathcal{L}$ is a linear combination of derivatives up to order $n$ applied to $\psi$, and each term in the sum over $i$ is a unique linear differential operator, $N_i$, of derivatives up to order $m_i$ that is applied to a nonlinear function $f_i$.

Under periodic boundary conditions, the semi-implicit Fourier spectral solver approximates the solution to $\psi$ by first applying the Fourier transform of Equation~(\ref{eq:spectralphasefield}):
\begin{align}
	\frac{\partial \hat{\psi}_{\vec{k}}}{\partial t} &= {L(k^n)}\hat{\psi}_{\vec{k}} + \sum_i{{N_i}(k^{m_i}) \hat{f}_{i}(\psi)_{\vec{k}}} 
	\,,
	\label{eq:phasefieldfourier}
\end{align}
where $\hat{\phantom{a}}$ indicates the Fourier transform of the respective term, $\hat{\psi}_{\vec{k}} = \hat{\psi}(\vec{k}, t)$, $\vec{k}$ is a vector in Fourier space and $k = |\vec{k}|$. Also, since $\vec{\nabla} \rightarrow i\vec{k}$ (and correspondingly $\nabla^2 \rightarrow -|k|^2$) under a Fourier transform in an infinite domain, the linear operator $\mathcal{L}$ becomes the function ${L}(k^n)$, a linear combination of the Fourier transformed derivatives, and likewise for $\mathcal{N}$. 

A difference scheme \cite{Provatas2010} to Equation~(\ref{eq:phasefieldfourier}) is determined by solving it as a linear ordinary differential equation and approximating to 1st order, yielding
\begin{align}
	\hat{\psi}(t + \Delta t) \approx A\hat{\psi}_{\vec{k}}(t) + B\sum_i N_i(k^{m_i}) \hat{f}_n(\psi)_{\vec{k}}\,,
	\label{eq:spectralscheme}
\end{align}
where
\begin{gather}
	A = e^{{L}(k^n)\Delta t} \quad \text{and} \quad B = \frac{e^{{L}(k^n)\Delta t} - 1}{{L}(k^n)}\,.
	\label{eq:spectraloperators}
\end{gather}

The spectral solver produces Equation~(\ref{eq:spectralscheme}) from any given equation of motion, a significant advantage that generalizes the spectral solver to a multitude of problems. This
demonstrates the adaptability of the solver and of the program design in general.
The spectral solver also computes the values of $A$ and $B$ \textit{a priori} to minimize the runtime.
The procedure is as follows:
\begin{enumerate}
	\item Split the equation into linear and nonlinear parts.
	Call the linear part $\mathcal{L}$ and the nonlinear part $\mathcal{N}$, analogous to the notation in Equation~(\ref{eq:spectralphasefield}).
	\item Further split the linear part by separating out terms which do not involve $\psi$, along with terms that cannot be expressed using a linear operator. 
	Call the expression formed by these terms $\mathcal{L}_*$ and the expression formed by all other terms $\mathcal{L}_\psi$. Thus, $\mathcal{L} = \mathcal{L}_\psi + \mathcal{L}_*$.
	\item Obtain ${L}_\psi$ by removing $\psi$ from terms in $\mathcal{L}_{\psi}$ and interchanging the derivative terms with the Fourier space transformed derivatives. Generate values for $A$ by evaluating ${L}_\psi$.
	\item Create the new expression ${{L}}_*$ by exchanging all order parameters in $\mathcal{L}_*$ with the Fourier transformed counterparts. 

	\item \label{step:make_D} Let ${{L}}_*$ be represented as the sum of its unique derivatives $d_n$ applied to expressions $e_n$, viz.: $$	{L}_* = \sum_n{d_n \cdot e_n}\,. $$ 
	Form the set $\mathbf{D}_* = \{(d_n, e_n) \mid n\}$.
	
	\item Apply Step~\ref{step:make_D} for the terms of $\mathcal{N}$, producing the list $\mathbf{D}_{\mathcal{N}}$. Form the set $\mathbf{D}_N = \{(d_n, \hat{e}_n) \mid (d_n, e_n) \in \mathbf{D}_{\mathcal{N}} \}$ where $\hat{\phantom{a}}$ denotes the Fourier transform of the respective term.
	
	\item 
	Define the following sets: 
	\begin{align}	
		\mathbf{D}_1 &= \{(d_n, e_n, e_m) \mid (d_n, e_n) \in \mathbf{D}_{*}, (d_m, e_m) \in \mathbf{D}_{{N}}, d_n = d_m\}  \,, \\
		\mathbf{D}_2 &= \{(d_n, e_n, 0) \mid (d_n, e_n) \in \mathbf{D}_{*}, (d_m, e_m) \in \mathbf{D}_{{N}}, d_n \not\in \{d\}_m\} \,, \\
		\mathbf{D}_3 &= \{(d_m, 0, e_m) \mid (d_m, e_m) \in \mathbf{D}_{{N}}, (d_n, e_n) \in \mathbf{D}_{*}, d_m \not\in \{d\}_n\}  \,.
	\end{align}
	Define $\mathbf{D} = \mathbf{D}_1 \,\cup\,\mathbf{D}_2\,\cup\,\mathbf{D}_3$. In other words, generate elements of $\mathbf{D}$ by pairing together the expressions in elements from $\mathbf{D}_{*}$ and $\mathbf{D}_{N}$ that match based on the derivatives $d_i$, and if there is no matching derivative in the other set, use 0 in place of the associated expression.
	
	\item Define two sequences ${B}_i = B\hat{d}_i$ where $B$ is as defined in Equation~(\ref{eq:spectraloperators}) and ${E}_i = (e_{n_i}, {e}_{m_i})$, using the elements: $$(d_i, e_{n_i}, e_{m_i}) \in \mathbf{D}.$$
	With respect to Equation~(\ref{eq:spectralscheme}), ${B}_i = B {N}_i(k)$ and $e_{n_i} + {e}_{m_i} = \hat{f}_i(\psi)_{\vec{k}}$.
	\item Return the set $\{A, \{{B}\}_i, \{{E}\}_i\}$.
\end{enumerate}

The scheme applied by the implemented spectral solver is then given by:
\begin{equation}
	\hat{\psi}^{n+1} = A\hat{\psi}^n(t) + \sum_i {B}_i \left( {E}^0_i + {E}^1_i \right)\,,
\end{equation}
where $\hat{\psi}^n$ is the approximate solution to $\hat{\psi}(\vec{k}, t)$  ($t = n\Delta t $ for time step $\Delta t$), ${E}^0_i$ and ${E}^1_i$ are the sets of the first and second elements of ${E}_i$, respectively, and subscript $i$ represents the indexed elements of their respective sets.

\section{Simulations and Verification} \label{sec:results}

Using the semi-implicit spectral solver, we performed simulations of the Allen--Cahn equation (Model A in the Hohenberg--Halperin classification \cite{Hohenberg1977}) describing a non-conserved order parameter $\psi = \psi(\vec{x}, t)$ \cite{Allen1975}, the Cahn--Hilliard equation (Model B in the Hohenberg--Halperin classification \cite{Hohenberg1977}) describing a conserved order parameter $m = m(\vec{x}, t)$, and a two order-parameter problem which couples a conserved order parameter $m = m(\vec{x}, t)$ with a non-conserved order parameter $\psi = \psi(\vec{x}, t)$ (Model C in the Hohenberg--Halperin classification \cite{Hohenberg1977}). The systems were simulated in both two and three dimensions. Since these systems are well-known and described in literature and since the main goal here is to demonstrate the software, we will not describe the above models in more detail but refer the reader to standard references such as the classic article by Hohenberg and Halperin \cite{Hohenberg1977} and the book by Provatas and Elder \cite{Provatas2010}. We also include a simulation of the phase-field crystal model of Elder et al.~\cite{Elder2002}.

The Allen--Cahn equation describes the dynamics of a non-conserved order parameter $\psi = \psi(\vec{x}, t)$ \cite{Hohenberg1977,Provatas2010, Allen1975} as 
\begin{equation}
        \frac{\partial \psi}{\partial t} = \nabla^2 \psi + c_1\psi -  c_3\psi^3\,.
        \label{eq:modelb}
\end{equation}
The Cahn--Hilliard equation, on the other hand, describes the phase separation dynamics of a conserved order parameter $m = m(\vec{x}, t)$ \cite{Hohenberg1977,Provatas2010,Cahn1958} as
\begin{equation}
    \frac{\partial m}{\partial t} = -\nabla^4 m - \nabla^2 \left(c_1m - c_2m^3 \right)\,,
    \label{eq:modela}
\end{equation}
In addition to these, we simulated a two order parameter problem which couples a conserved order parameter $m = m(\vec{x}, t)$ with a non-conserved order parameter $\psi = \psi(\vec{x}, t)$ (Model C~\cite{Hohenberg1977}) through a nonlinear term in the free energy functional:
\begin{align}
    \dfrac{\partial \psi}{\partial t} &= 
    \nabla^2 \psi +c_1\psi - c_2\psi^3 + 2c_5 \psi m\\
    \dfrac{\partial m}{\partial t} &= -\nabla^4 m - \nabla^2 \left(c_3 m - c_4 m^3 + c_5 \psi^2\right).
    \label{eq:modelc}
\end{align}
This model has been used to describe eutectic growth~\cite{Elder1994}.

The phase-field crystal model was developed to include periodic structure to the standard phase-field free energy functional in order to represent elastic and plastic interactions in a crystal \cite{Elder2002, Elder_2004}.
For a conserved density field $n=n(\vec{x}, t)$, the dynamical equation of a phase-field crystal model is given by:
\begin{equation}
	\frac{\partial n}{\partial t} = \nabla^2\left( n^2 + n^3 + \left((q_0 + \nabla^2)^2 - \varepsilon\right)n \right). \label{eq:pfc}
\end{equation}

Coarsening of the phase-field at any point during the transition can be quantified by the radial average of the static structure factor, $S(k)$. The static structure factor, $S(\vec{k})$, measures incident scattering in a solid \cite{Lovesey1984, Squires1978}. In the Born approximation for periodic systems, $S(\vec{k}) = | \hat{\rho}_{\vec{k}} |^2$ \cite{Lindgard1994}. The term $\hat{\rho}_{\vec{k}}$ is the Fourier transform of $\rho(\vec{r})$, the particle occupancy at the position $\vec{r}$ in the lattice, which is 1 in the solid phase and 0 elsewhere. Correspondingly, for computing the structure factor of the continuous order parameter field $\psi(\vec{r})$, the positive phase is chosen to represent solidification, that is, $\rho = 1$ when $\psi > 0$ and $\rho = 0$ otherwise. 

For both the conserved and non-conserved phase-field models, the radial average of the structure factor should correspond to Porod's law~\cite{Bray_2002}:
\begin{equation}
    S(k) \sim (Lk^{d+1})^{-1}\,,
    \label{eq:porodslaw}
\end{equation}
where $L$ is the size of the system and $d$ is the dimension. 
We use this to establish that \SymPhas{} generates correct solutions by validating that the relationship holds for $S(k)$ measured from simulations of the Allen--Cahn and Cahn--Hilliard models, representing the non-conserved and conserved dynamics, respectively~\cite{Puri1997}. We proceed by computing $S(k)$ from the average of 10 independent simulations of these models in both 2D and 3D and verifying that the scaling is consistent to Porod's law, Equation~(\ref{eq:porodslaw}), for the corresponding dimension. 

The implementations of the simulated models with the \SymPhas{} model definitions macros are provided in Figure~\ref{fig:models_abc_definitions}. As the figure shows, models are defined in a compact and intuitive way.

\begin{figure}
    \centering
\begin{minipage}{0.48\textwidth}
    %
    \begin{subfigure}{\textwidth}
    \centering
    \begingroup
    \inputencoding{cp1252}%
    \makeatletter
    \providecommand\color[2][]{%
    	\GenericError{(gnuplot) \space\space\space\@spaces}{%
    		Package color not loaded in conjunction with
    		terminal option `colourtext'%
    	}{See the gnuplot documentation for explanation.%
    	}{Either use 'blacktext' in gnuplot or load the package
    		color.sty in LaTeX.}%
    	\renewcommand\color[2][]{}%
    }%
    \providecommand\includegraphics[2][]{%
    	\GenericError{(gnuplot) \space\space\space\@spaces}{%
    		Package graphicx or graphics not loaded%
    	}{See the gnuplot documentation for explanation.%
    	}{The gnuplot epslatex terminal needs graphicx.sty or graphics.sty.}%
    	\renewcommand\includegraphics[2][]{}%
    }%
    \providecommand\rotatebox[2]{#2}%
    \@ifundefined{ifGPcolor}{%
    	\newif\ifGPcolor
    	\GPcolortrue
    }{}%
    \@ifundefined{ifGPblacktext}{%
    	\newif\ifGPblacktext
    	\GPblacktexttrue
    }{}%
    \let\gplgaddtomacro\g@addto@macro
    \gdef\gplbacktext{}%
    \gdef\gplfronttext{}%
    \makeatother
    \ifGPblacktext
    \def\colorrgb#1{}%
    \def\colorgray#1{}%
    \else
    \ifGPcolor
    \def\colorrgb#1{\color[rgb]{#1}}%
    \def\colorgray#1{\color[gray]{#1}}%
    \expandafter\def\csname LTw\endcsname{\color{white}}%
    \expandafter\def\csname LTb\endcsname{\color{black}}%
    \expandafter\def\csname LTa\endcsname{\color{black}}%
    \expandafter\def\csname LT0\endcsname{\color[rgb]{1,0,0}}%
    \expandafter\def\csname LT1\endcsname{\color[rgb]{0,1,0}}%
    \expandafter\def\csname LT2\endcsname{\color[rgb]{0,0,1}}%
    \expandafter\def\csname LT3\endcsname{\color[rgb]{1,0,1}}%
    \expandafter\def\csname LT4\endcsname{\color[rgb]{0,1,1}}%
    \expandafter\def\csname LT5\endcsname{\color[rgb]{1,1,0}}%
    \expandafter\def\csname LT6\endcsname{\color[rgb]{0,0,0}}%
    \expandafter\def\csname LT7\endcsname{\color[rgb]{1,0.3,0}}%
    \expandafter\def\csname LT8\endcsname{\color[rgb]{0.5,0.5,0.5}}%
    \else
    \def\colorrgb#1{\color{black}}%
    \def\colorgray#1{\color[gray]{#1}}%
    \expandafter\def\csname LTw\endcsname{\color{white}}%
    \expandafter\def\csname LTb\endcsname{\color{black}}%
    \expandafter\def\csname LTa\endcsname{\color{black}}%
    \expandafter\def\csname LT0\endcsname{\color{black}}%
    \expandafter\def\csname LT1\endcsname{\color{black}}%
    \expandafter\def\csname LT2\endcsname{\color{black}}%
    \expandafter\def\csname LT3\endcsname{\color{black}}%
    \expandafter\def\csname LT4\endcsname{\color{black}}%
    \expandafter\def\csname LT5\endcsname{\color{black}}%
    \expandafter\def\csname LT6\endcsname{\color{black}}%
    \expandafter\def\csname LT7\endcsname{\color{black}}%
    \expandafter\def\csname LT8\endcsname{\color{black}}%
    \fi
    \fi
    \setlength{\unitlength}{0.0500bp}%
    \ifx\gptboxheight\undefined%
    \newlength{\gptboxheight}%
    \newlength{\gptboxwidth}%
    \newsavebox{\gptboxtext}%
    \fi%
    \setlength{\fboxrule}{0.5pt}%
    \setlength{\fboxsep}{1pt}%
    \begin{picture}(5182.00,1872.00)%
    	\gplgaddtomacro\gplbacktext{%
    	}%
    	\gplgaddtomacro\gplfronttext{%
    		\csname LTb\endcsname
    		\put(1714,1924){\makebox(0,0)[r]{\strut{}index $500$}}%
    	}%
    	\gplgaddtomacro\gplbacktext{%
    	}%
    	\gplgaddtomacro\gplfronttext{%
    		\csname LTb\endcsname
    		\put(3341,1924){\makebox(0,0)[r]{\strut{}index $2,000$}}%
    	}%
    	\gplgaddtomacro\gplbacktext{%
    	}%
    	\gplgaddtomacro\gplfronttext{%
    		\csname LTb\endcsname
    		\put(4963,1924){\makebox(0,0)[r]{\strut{}index $20,000$}}%
    		\csname LTb\endcsname
    		\put(218,0){\makebox(0,0){\strut{}$-1$}}%
    		\put(1404,0){\makebox(0,0){\strut{}$-0.5$}}%
    		\put(2590,0){\makebox(0,0){\strut{}$0$}}%
    		\put(3776,0){\makebox(0,0){\strut{}$0.5$}}%
    		\put(4963,0){\makebox(0,0){\strut{}$1$}}%
    	}%
    	\gplbacktext
    	\put(0,0){\includegraphics{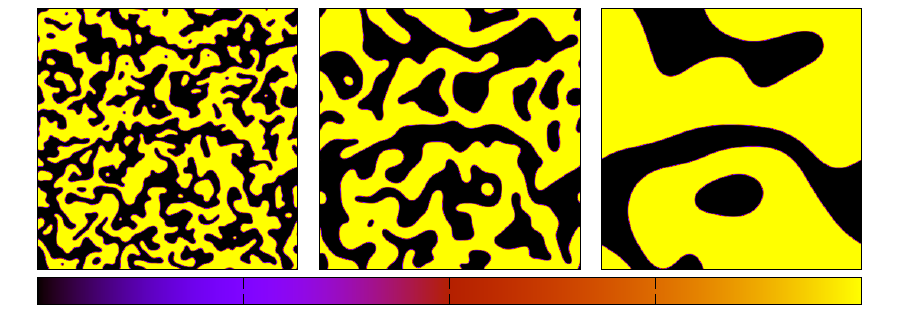}}%
    	\gplfronttext
    \end{picture}%
    \endgroup
    
    \caption{}
    \label{fig:modelab:a:2d}
    \end{subfigure}
    \begin{subfigure}{\textwidth}
    \centering
    \includegraphics[width=0.33\textwidth]{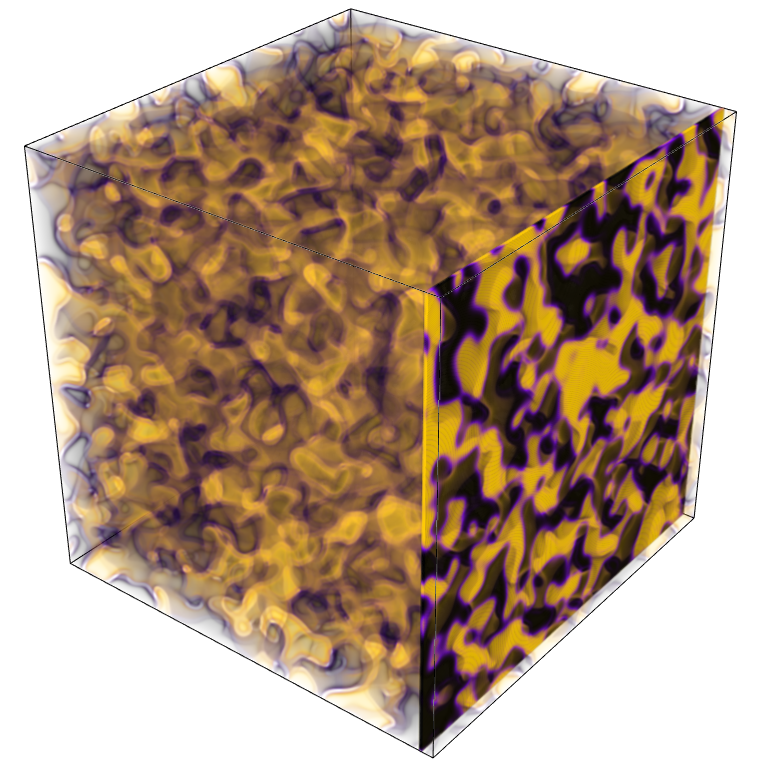}\hfill%
    \includegraphics[width=0.33\textwidth]{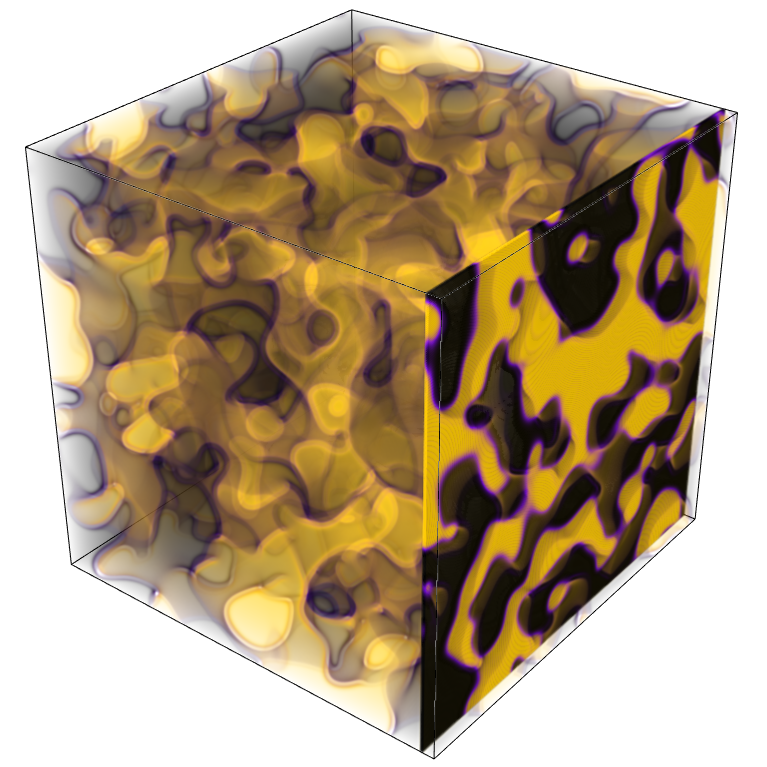}\hfill%
    \includegraphics[width=0.33\textwidth]{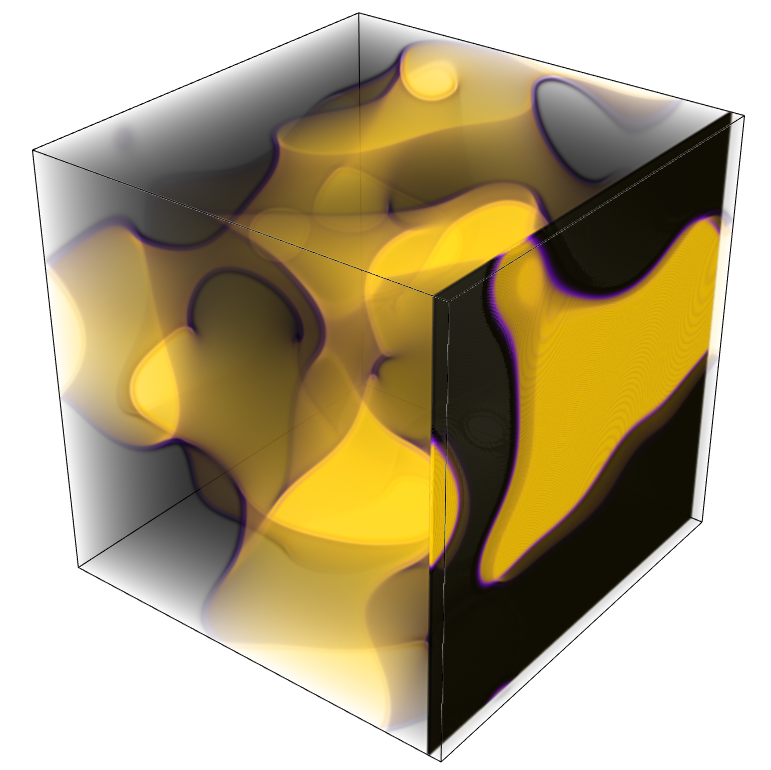}
    \caption{}
    \label{fig:modelab:a:cs}
    \end{subfigure}\vspace{10pt}
    \begin{subfigure}{\textwidth}
    \centering
    
    \begingroup
    \inputencoding{cp1252}%
    \makeatletter
    \providecommand\color[2][]{%
    	\GenericError{(gnuplot) \space\space\space\@spaces}{%
    		Package color not loaded in conjunction with
    		terminal option `colourtext'%
    	}{See the gnuplot documentation for explanation.%
    	}{Either use 'blacktext' in gnuplot or load the package
    		color.sty in LaTeX.}%
    	\renewcommand\color[2][]{}%
    }%
    \providecommand\includegraphics[2][]{%
    	\GenericError{(gnuplot) \space\space\space\@spaces}{%
    		Package graphicx or graphics not loaded%
    	}{See the gnuplot documentation for explanation.%
    	}{The gnuplot epslatex terminal needs graphicx.sty or graphics.sty.}%
    	\renewcommand\includegraphics[2][]{}%
    }%
    \providecommand\rotatebox[2]{#2}%
    \@ifundefined{ifGPcolor}{%
    	\newif\ifGPcolor
    	\GPcolortrue
    }{}%
    \@ifundefined{ifGPblacktext}{%
    	\newif\ifGPblacktext
    	\GPblacktexttrue
    }{}%
    \let\gplgaddtomacro\g@addto@macro
    \gdef\gplbacktext{}%
    \gdef\gplfronttext{}%
    \makeatother
    \ifGPblacktext
    \def\colorrgb#1{}%
    \def\colorgray#1{}%
    \else
    \ifGPcolor
    \def\colorrgb#1{\color[rgb]{#1}}%
    \def\colorgray#1{\color[gray]{#1}}%
    \expandafter\def\csname LTw\endcsname{\color{white}}%
    \expandafter\def\csname LTb\endcsname{\color{black}}%
    \expandafter\def\csname LTa\endcsname{\color{black}}%
    \expandafter\def\csname LT0\endcsname{\color[rgb]{1,0,0}}%
    \expandafter\def\csname LT1\endcsname{\color[rgb]{0,1,0}}%
    \expandafter\def\csname LT2\endcsname{\color[rgb]{0,0,1}}%
    \expandafter\def\csname LT3\endcsname{\color[rgb]{1,0,1}}%
    \expandafter\def\csname LT4\endcsname{\color[rgb]{0,1,1}}%
    \expandafter\def\csname LT5\endcsname{\color[rgb]{1,1,0}}%
    \expandafter\def\csname LT6\endcsname{\color[rgb]{0,0,0}}%
    \expandafter\def\csname LT7\endcsname{\color[rgb]{1,0.3,0}}%
    \expandafter\def\csname LT8\endcsname{\color[rgb]{0.5,0.5,0.5}}%
    \else
    \def\colorrgb#1{\color{black}}%
    \def\colorgray#1{\color[gray]{#1}}%
    \expandafter\def\csname LTw\endcsname{\color{white}}%
    \expandafter\def\csname LTb\endcsname{\color{black}}%
    \expandafter\def\csname LTa\endcsname{\color{black}}%
    \expandafter\def\csname LT0\endcsname{\color{black}}%
    \expandafter\def\csname LT1\endcsname{\color{black}}%
    \expandafter\def\csname LT2\endcsname{\color{black}}%
    \expandafter\def\csname LT3\endcsname{\color{black}}%
    \expandafter\def\csname LT4\endcsname{\color{black}}%
    \expandafter\def\csname LT5\endcsname{\color{black}}%
    \expandafter\def\csname LT6\endcsname{\color{black}}%
    \expandafter\def\csname LT7\endcsname{\color{black}}%
    \expandafter\def\csname LT8\endcsname{\color{black}}%
    \fi
    \fi
    \setlength{\unitlength}{0.0500bp}%
    \ifx\gptboxheight\undefined%
    \newlength{\gptboxheight}%
    \newlength{\gptboxwidth}%
    \newsavebox{\gptboxtext}%
    \fi%
    \setlength{\fboxrule}{0.5pt}%
    \setlength{\fboxsep}{1pt}%
    \begin{picture}(5182.00,1872.00)%
    	\gplgaddtomacro\gplbacktext{%
    	}%
    	\gplgaddtomacro\gplfronttext{%
    		\csname LTb\endcsname
    		\put(1714,1924){\makebox(0,0)[r]{\strut{}index $50$}}%
    	}%
    	\gplgaddtomacro\gplbacktext{%
    	}%
    	\gplgaddtomacro\gplfronttext{%
    		\csname LTb\endcsname
    		\put(3341,1924){\makebox(0,0)[r]{\strut{}index $200$}}%
    	}%
    	\gplgaddtomacro\gplbacktext{%
    	}%
    	\gplgaddtomacro\gplfronttext{%
    		\csname LTb\endcsname
    		\put(4963,1924){\makebox(0,0)[r]{\strut{}index $2,000$}}%
    		\csname LTb\endcsname
    		\put(218,0){\makebox(0,0){\strut{}$-1$}}%
    		\put(1404,0){\makebox(0,0){\strut{}$-0.5$}}%
    		\put(2590,0){\makebox(0,0){\strut{}$0$}}%
    		\put(3776,0){\makebox(0,0){\strut{}$0.5$}}%
    		\put(4963,0){\makebox(0,0){\strut{}$1$}}%
    	}%
    	\gplbacktext
    	\put(0,0){\includegraphics{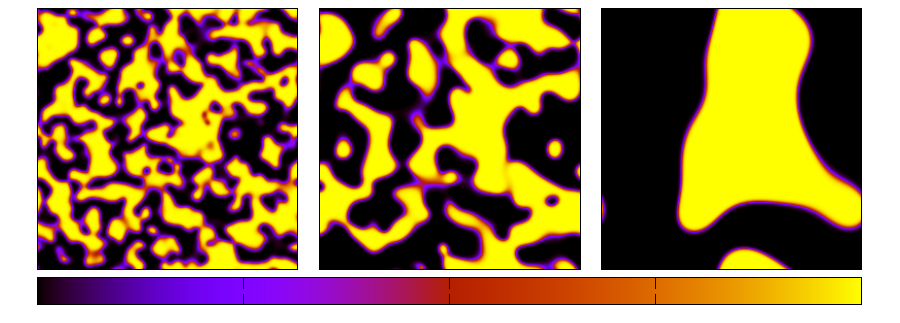}}%
    	\gplfronttext
    \end{picture}%
    \endgroup

    \caption{}
    \label{fig:modelab:a:3d}
    \end{subfigure}
    \caption{Snapshots from simulations of a 2D ($1024\times1024$) and 3D ($256 \!\times256 \times \!256$) Allen--Cahn model \cite{Allen1975}:
    (\subref{fig:modelab:a:2d}) the 2D system is shown at three intervals at solution index 500, 2,000 and 20,000.  (\subref{fig:modelab:a:cs}) the 3D system is visualized using VTK \cite{vtk} at solution index 50, 200 and 2,000 with a cross-section highlighted for visibility, where (\subref{fig:modelab:a:3d}) shows the cross sections.
    The simulations use a time step of $\Delta t = 0.25$ and are initially seeded with random values between -1 and 1.
    }
    \label{fig:modelab:a}
\end{minipage}\hfill%
\begin{minipage}{0.48\textwidth}
    %
    \centering
    \begin{subfigure}{\textwidth}
    \centering
    
    \begingroup
    \inputencoding{cp1252}%
    \makeatletter
    \providecommand\color[2][]{%
    	\GenericError{(gnuplot) \space\space\space\@spaces}{%
    		Package color not loaded in conjunction with
    		terminal option `colourtext'%
    	}{See the gnuplot documentation for explanation.%
    	}{Either use 'blacktext' in gnuplot or load the package
    		color.sty in LaTeX.}%
    	\renewcommand\color[2][]{}%
    }%
    \providecommand\includegraphics[2][]{%
    	\GenericError{(gnuplot) \space\space\space\@spaces}{%
    		Package graphicx or graphics not loaded%
    	}{See the gnuplot documentation for explanation.%
    	}{The gnuplot epslatex terminal needs graphicx.sty or graphics.sty.}%
    	\renewcommand\includegraphics[2][]{}%
    }%
    \providecommand\rotatebox[2]{#2}%
    \@ifundefined{ifGPcolor}{%
    	\newif\ifGPcolor
    	\GPcolortrue
    }{}%
    \@ifundefined{ifGPblacktext}{%
    	\newif\ifGPblacktext
    	\GPblacktexttrue
    }{}%
    \let\gplgaddtomacro\g@addto@macro
    \gdef\gplbacktext{}%
    \gdef\gplfronttext{}%
    \makeatother
    \ifGPblacktext
    \def\colorrgb#1{}%
    \def\colorgray#1{}%
    \else
    \ifGPcolor
    \def\colorrgb#1{\color[rgb]{#1}}%
    \def\colorgray#1{\color[gray]{#1}}%
    \expandafter\def\csname LTw\endcsname{\color{white}}%
    \expandafter\def\csname LTb\endcsname{\color{black}}%
    \expandafter\def\csname LTa\endcsname{\color{black}}%
    \expandafter\def\csname LT0\endcsname{\color[rgb]{1,0,0}}%
    \expandafter\def\csname LT1\endcsname{\color[rgb]{0,1,0}}%
    \expandafter\def\csname LT2\endcsname{\color[rgb]{0,0,1}}%
    \expandafter\def\csname LT3\endcsname{\color[rgb]{1,0,1}}%
    \expandafter\def\csname LT4\endcsname{\color[rgb]{0,1,1}}%
    \expandafter\def\csname LT5\endcsname{\color[rgb]{1,1,0}}%
    \expandafter\def\csname LT6\endcsname{\color[rgb]{0,0,0}}%
    \expandafter\def\csname LT7\endcsname{\color[rgb]{1,0.3,0}}%
    \expandafter\def\csname LT8\endcsname{\color[rgb]{0.5,0.5,0.5}}%
    \else
    \def\colorrgb#1{\color{black}}%
    \def\colorgray#1{\color[gray]{#1}}%
    \expandafter\def\csname LTw\endcsname{\color{white}}%
    \expandafter\def\csname LTb\endcsname{\color{black}}%
    \expandafter\def\csname LTa\endcsname{\color{black}}%
    \expandafter\def\csname LT0\endcsname{\color{black}}%
    \expandafter\def\csname LT1\endcsname{\color{black}}%
    \expandafter\def\csname LT2\endcsname{\color{black}}%
    \expandafter\def\csname LT3\endcsname{\color{black}}%
    \expandafter\def\csname LT4\endcsname{\color{black}}%
    \expandafter\def\csname LT5\endcsname{\color{black}}%
    \expandafter\def\csname LT6\endcsname{\color{black}}%
    \expandafter\def\csname LT7\endcsname{\color{black}}%
    \expandafter\def\csname LT8\endcsname{\color{black}}%
    \fi
    \fi
    \setlength{\unitlength}{0.0500bp}%
    \ifx\gptboxheight\undefined%
    \newlength{\gptboxheight}%
    \newlength{\gptboxwidth}%
    \newsavebox{\gptboxtext}%
    \fi%
    \setlength{\fboxrule}{0.5pt}%
    \setlength{\fboxsep}{1pt}%
    \begin{picture}(5182.00,1872.00)%
    	\gplgaddtomacro\gplbacktext{%
    	}%
    	\gplgaddtomacro\gplfronttext{%
    		\csname LTb\endcsname
    		\put(1714,1924){\makebox(0,0)[r]{\strut{}index $5,000$}}%
    	}%
    	\gplgaddtomacro\gplbacktext{%
    	}%
    	\gplgaddtomacro\gplfronttext{%
    		\csname LTb\endcsname
    		\put(3341,1924){\makebox(0,0)[r]{\strut{}index $20,000$}}%
    	}%
    	\gplgaddtomacro\gplbacktext{%
    	}%
    	\gplgaddtomacro\gplfronttext{%
    		\csname LTb\endcsname
    		\put(4963,1924){\makebox(0,0)[r]{\strut{}index $200,000$}}%
    		\csname LTb\endcsname
    		\put(218,0){\makebox(0,0){\strut{}$-1$}}%
    		\put(1404,0){\makebox(0,0){\strut{}$-0.5$}}%
    		\put(2590,0){\makebox(0,0){\strut{}$0$}}%
    		\put(3776,0){\makebox(0,0){\strut{}$0.5$}}%
    		\put(4963,0){\makebox(0,0){\strut{}$1$}}%
    	}%
    	\gplbacktext
    	\put(0,0){\includegraphics{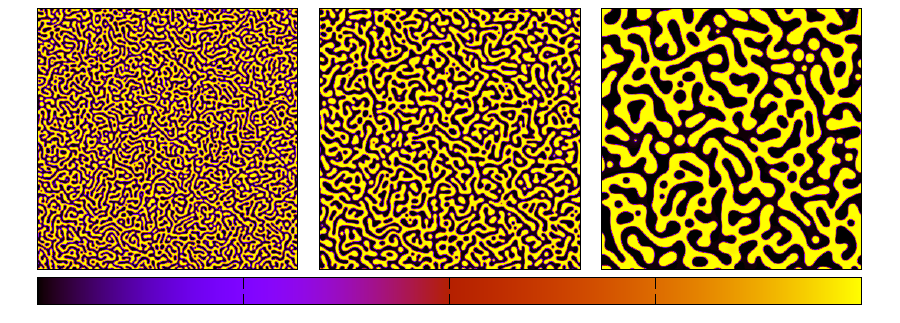}}%
    	\gplfronttext
    \end{picture}%
    \endgroup

    \caption{}
    \label{fig:modelab:b:2d}
    \end{subfigure}
    \begin{subfigure}{\textwidth}
    \centering
    \includegraphics[width=0.33\textwidth]{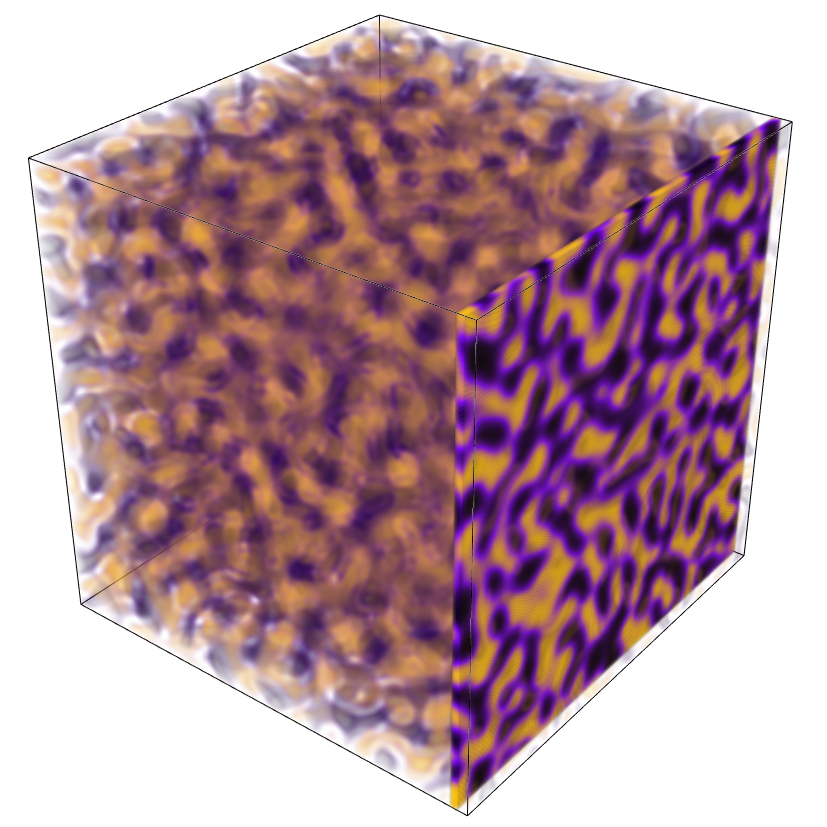}\hfill%
    \includegraphics[width=0.33\textwidth]{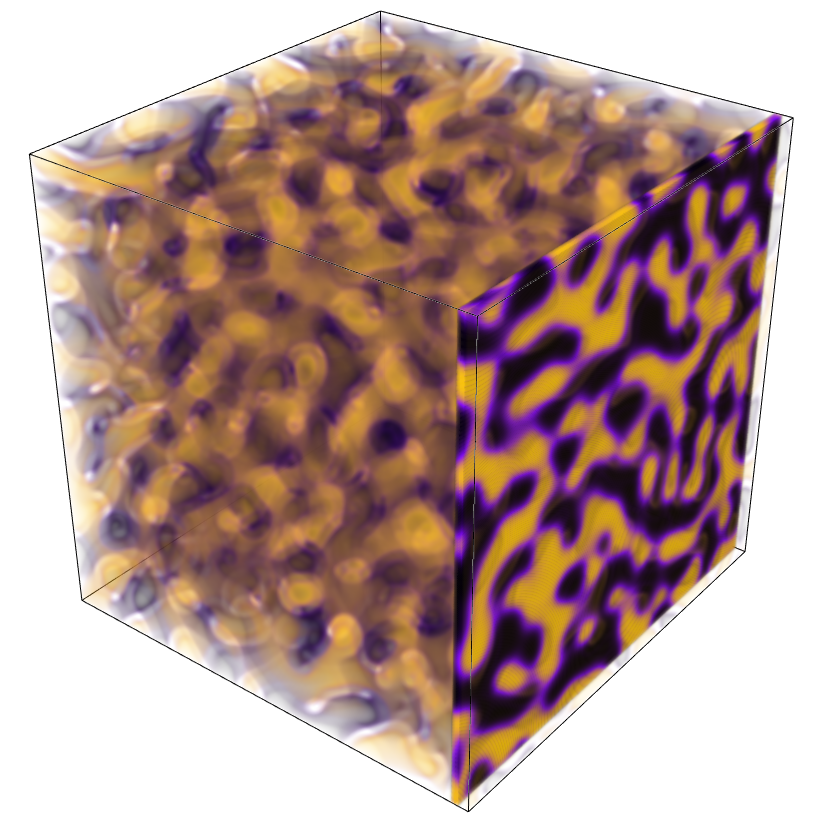}\hfill%
    \includegraphics[width=0.33\textwidth]{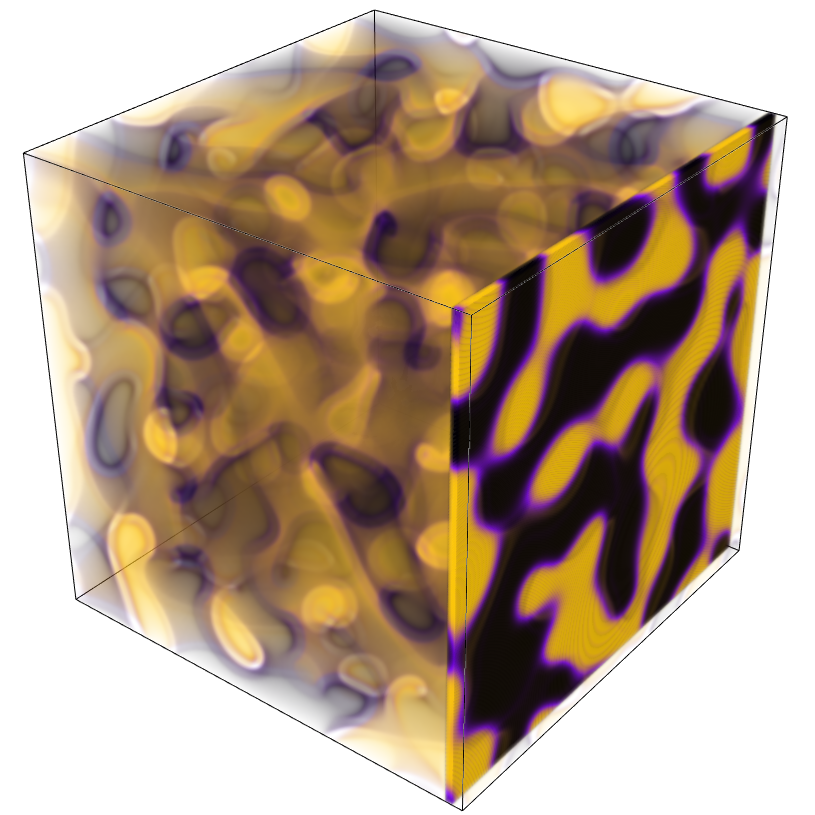}
    \caption{}
    \label{fig:modelab:b:cs}
    \end{subfigure}\vspace{10pt}
    \begin{subfigure}{\textwidth}
    \centering
    
    \begingroup
    \inputencoding{cp1252}%
    \makeatletter
    \providecommand\color[2][]{%
    	\GenericError{(gnuplot) \space\space\space\@spaces}{%
    		Package color not loaded in conjunction with
    		terminal option `colourtext'%
    	}{See the gnuplot documentation for explanation.%
    	}{Either use 'blacktext' in gnuplot or load the package
    		color.sty in LaTeX.}%
    	\renewcommand\color[2][]{}%
    }%
    \providecommand\includegraphics[2][]{%
    	\GenericError{(gnuplot) \space\space\space\@spaces}{%
    		Package graphicx or graphics not loaded%
    	}{See the gnuplot documentation for explanation.%
    	}{The gnuplot epslatex terminal needs graphicx.sty or graphics.sty.}%
    	\renewcommand\includegraphics[2][]{}%
    }%
    \providecommand\rotatebox[2]{#2}%
    \@ifundefined{ifGPcolor}{%
    	\newif\ifGPcolor
    	\GPcolortrue
    }{}%
    \@ifundefined{ifGPblacktext}{%
    	\newif\ifGPblacktext
    	\GPblacktexttrue
    }{}%
    \let\gplgaddtomacro\g@addto@macro
    \gdef\gplbacktext{}%
    \gdef\gplfronttext{}%
    \makeatother
    \ifGPblacktext
    \def\colorrgb#1{}%
    \def\colorgray#1{}%
    \else
    \ifGPcolor
    \def\colorrgb#1{\color[rgb]{#1}}%
    \def\colorgray#1{\color[gray]{#1}}%
    \expandafter\def\csname LTw\endcsname{\color{white}}%
    \expandafter\def\csname LTb\endcsname{\color{black}}%
    \expandafter\def\csname LTa\endcsname{\color{black}}%
    \expandafter\def\csname LT0\endcsname{\color[rgb]{1,0,0}}%
    \expandafter\def\csname LT1\endcsname{\color[rgb]{0,1,0}}%
    \expandafter\def\csname LT2\endcsname{\color[rgb]{0,0,1}}%
    \expandafter\def\csname LT3\endcsname{\color[rgb]{1,0,1}}%
    \expandafter\def\csname LT4\endcsname{\color[rgb]{0,1,1}}%
    \expandafter\def\csname LT5\endcsname{\color[rgb]{1,1,0}}%
    \expandafter\def\csname LT6\endcsname{\color[rgb]{0,0,0}}%
    \expandafter\def\csname LT7\endcsname{\color[rgb]{1,0.3,0}}%
    \expandafter\def\csname LT8\endcsname{\color[rgb]{0.5,0.5,0.5}}%
    \else
    \def\colorrgb#1{\color{black}}%
    \def\colorgray#1{\color[gray]{#1}}%
    \expandafter\def\csname LTw\endcsname{\color{white}}%
    \expandafter\def\csname LTb\endcsname{\color{black}}%
    \expandafter\def\csname LTa\endcsname{\color{black}}%
    \expandafter\def\csname LT0\endcsname{\color{black}}%
    \expandafter\def\csname LT1\endcsname{\color{black}}%
    \expandafter\def\csname LT2\endcsname{\color{black}}%
    \expandafter\def\csname LT3\endcsname{\color{black}}%
    \expandafter\def\csname LT4\endcsname{\color{black}}%
    \expandafter\def\csname LT5\endcsname{\color{black}}%
    \expandafter\def\csname LT6\endcsname{\color{black}}%
    \expandafter\def\csname LT7\endcsname{\color{black}}%
    \expandafter\def\csname LT8\endcsname{\color{black}}%
    \fi
    \fi
    \setlength{\unitlength}{0.0500bp}%
    \ifx\gptboxheight\undefined%
    \newlength{\gptboxheight}%
    \newlength{\gptboxwidth}%
    \newsavebox{\gptboxtext}%
    \fi%
    \setlength{\fboxrule}{0.5pt}%
    \setlength{\fboxsep}{1pt}%
    \begin{picture}(5182.00,1872.00)%
    	\gplgaddtomacro\gplbacktext{%
    	}%
    	\gplgaddtomacro\gplfronttext{%
    		\csname LTb\endcsname
    		\put(1714,1924){\makebox(0,0)[r]{\strut{}index $500$}}%
    	}%
    	\gplgaddtomacro\gplbacktext{%
    	}%
    	\gplgaddtomacro\gplfronttext{%
    		\csname LTb\endcsname
    		\put(3341,1924){\makebox(0,0)[r]{\strut{}index $2,000$}}%
    	}%
    	\gplgaddtomacro\gplbacktext{%
    	}%
    	\gplgaddtomacro\gplfronttext{%
    		\csname LTb\endcsname
    		\put(4963,1924){\makebox(0,0)[r]{\strut{}index $20,000$}}%
    		\csname LTb\endcsname
    		\put(218,0){\makebox(0,0){\strut{}$-1$}}%
    		\put(1404,0){\makebox(0,0){\strut{}$-0.5$}}%
    		\put(2590,0){\makebox(0,0){\strut{}$0$}}%
    		\put(3776,0){\makebox(0,0){\strut{}$0.5$}}%
    		\put(4963,0){\makebox(0,0){\strut{}$1$}}%
    	}%
    	\gplbacktext
    	\put(0,0){\includegraphics{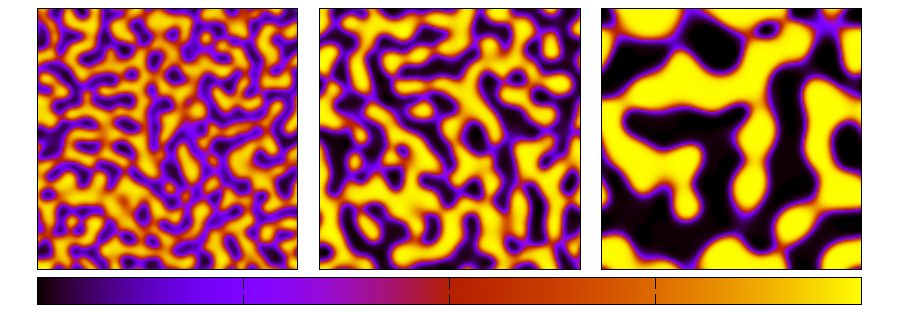}}%
    	\gplfronttext
    \end{picture}%
    \endgroup

    \caption{}
    \label{fig:modelab:b:3d}
    \end{subfigure}
    \caption{Snapshots from simulations of a 2D ($1024\times1024$) and 3D ($128  \! \times \! 128 \! \times \!128$) Cahn--Hilliard model \cite{Cahn1958}: (\subref{fig:modelab:b:2d}) the 2D system is shown at three intervals at solution index 5,000, 20,000 and 200,000; and (\subref{fig:modelab:a:cs}) the 3D system is visualized using VTK \cite{vtk} at solution index 2,500, 5,000 and 20,000 with a cross-section highlighted for visibility, where (\subref{fig:modelab:a:3d}) shows the cross sections.
    The simulations use a time step of $\Delta t = 0.05$, and are initially seeded with random values between -1 and 1.}
    \label{fig:modelab:b}
\end{minipage}
\end{figure}

\begin{figure}
    \centering
    \begin{subfigure}{.5\textwidth}
    \centering
    
    \begingroup
    \inputencoding{cp1252}%
    \makeatletter
    \providecommand\color[2][]{%
    	\GenericError{(gnuplot) \space\space\space\@spaces}{%
    		Package color not loaded in conjunction with
    		terminal option `colourtext'%
    	}{See the gnuplot documentation for explanation.%
    	}{Either use 'blacktext' in gnuplot or load the package
    		color.sty in LaTeX.}%
    	\renewcommand\color[2][]{}%
    }%
    \providecommand\includegraphics[2][]{%
    	\GenericError{(gnuplot) \space\space\space\@spaces}{%
    		Package graphicx or graphics not loaded%
    	}{See the gnuplot documentation for explanation.%
    	}{The gnuplot epslatex terminal needs graphicx.sty or graphics.sty.}%
    	\renewcommand\includegraphics[2][]{}%
    }%
    \providecommand\rotatebox[2]{#2}%
    \@ifundefined{ifGPcolor}{%
    	\newif\ifGPcolor
    	\GPcolortrue
    }{}%
    \@ifundefined{ifGPblacktext}{%
    	\newif\ifGPblacktext
    	\GPblacktexttrue
    }{}%
    \let\gplgaddtomacro\g@addto@macro
    \gdef\gplbacktext{}%
    \gdef\gplfronttext{}%
    \makeatother
    \ifGPblacktext
    \def\colorrgb#1{}%
    \def\colorgray#1{}%
    \else
    \ifGPcolor
    \def\colorrgb#1{\color[rgb]{#1}}%
    \def\colorgray#1{\color[gray]{#1}}%
    \expandafter\def\csname LTw\endcsname{\color{white}}%
    \expandafter\def\csname LTb\endcsname{\color{black}}%
    \expandafter\def\csname LTa\endcsname{\color{black}}%
    \expandafter\def\csname LT0\endcsname{\color[rgb]{1,0,0}}%
    \expandafter\def\csname LT1\endcsname{\color[rgb]{0,1,0}}%
    \expandafter\def\csname LT2\endcsname{\color[rgb]{0,0,1}}%
    \expandafter\def\csname LT3\endcsname{\color[rgb]{1,0,1}}%
    \expandafter\def\csname LT4\endcsname{\color[rgb]{0,1,1}}%
    \expandafter\def\csname LT5\endcsname{\color[rgb]{1,1,0}}%
    \expandafter\def\csname LT6\endcsname{\color[rgb]{0,0,0}}%
    \expandafter\def\csname LT7\endcsname{\color[rgb]{1,0.3,0}}%
    \expandafter\def\csname LT8\endcsname{\color[rgb]{0.5,0.5,0.5}}%
    \else
    \def\colorrgb#1{\color{black}}%
    \def\colorgray#1{\color[gray]{#1}}%
    \expandafter\def\csname LTw\endcsname{\color{white}}%
    \expandafter\def\csname LTb\endcsname{\color{black}}%
    \expandafter\def\csname LTa\endcsname{\color{black}}%
    \expandafter\def\csname LT0\endcsname{\color{black}}%
    \expandafter\def\csname LT1\endcsname{\color{black}}%
    \expandafter\def\csname LT2\endcsname{\color{black}}%
    \expandafter\def\csname LT3\endcsname{\color{black}}%
    \expandafter\def\csname LT4\endcsname{\color{black}}%
    \expandafter\def\csname LT5\endcsname{\color{black}}%
    \expandafter\def\csname LT6\endcsname{\color{black}}%
    \expandafter\def\csname LT7\endcsname{\color{black}}%
    \expandafter\def\csname LT8\endcsname{\color{black}}%
    \fi
    \fi
    \setlength{\unitlength}{0.0500bp}%
    \ifx\gptboxheight\undefined%
    \newlength{\gptboxheight}%
    \newlength{\gptboxwidth}%
    \newsavebox{\gptboxtext}%
    \fi%
    \setlength{\fboxrule}{0.5pt}%
    \setlength{\fboxsep}{1pt}%
    \begin{picture}(4320.00,5760.00)%
    	\gplgaddtomacro\gplbacktext{%
    		\csname LTb\endcsname
    		\put(682,704){\makebox(0,0)[r]{\strut{}-4}}%
    		\put(682,1255){\makebox(0,0)[r]{\strut{}-3}}%
    		\put(682,1806){\makebox(0,0)[r]{\strut{}-2}}%
    		\put(682,2356){\makebox(0,0)[r]{\strut{}-1}}%
    		\put(682,2907){\makebox(0,0)[r]{\strut{}0}}%
    		\put(682,3458){\makebox(0,0)[r]{\strut{}1}}%
    		\put(682,4009){\makebox(0,0)[r]{\strut{}2}}%
    		\put(682,4560){\makebox(0,0)[r]{\strut{}3}}%
    		\put(682,5110){\makebox(0,0)[r]{\strut{}4}}%
    		\put(1054,484){\makebox(0,0){\strut{}$0.01$}}%
    		\put(2138,484){\makebox(0,0){\strut{}$0.1$}}%
    		\put(3221,484){\makebox(0,0){\strut{}$1$}}%
    	}%
    	\gplgaddtomacro\gplfronttext{%
    		\csname LTb\endcsname
    		\put(198,3121){\rotatebox{-270}{\makebox(0,0){\strut{}$S(k)$}}}%
    		\put(2368,154){\makebox(0,0){\strut{}$k$}}%
    		\put(2835,5333){\makebox(0,0){\strut{}Model A, 2D}}%
    		\csname LTb\endcsname
    		\put(3200,4170){\makebox(0,0)[r]{\strut{}500}}%
    		\csname LTb\endcsname
    		\put(3200,4456){\makebox(0,0)[r]{\strut{}2,000}}%
    		\csname LTb\endcsname
    		\put(3200,4742){\makebox(0,0)[r]{\strut{}20,000}}%
    		\csname LTb\endcsname
    		\put(3200,5028){\makebox(0,0)[r]{\strut{}$S(k) \sim k^{-3}$}}%
    	}%
    	\gplbacktext
    	\put(0,0){\includegraphics{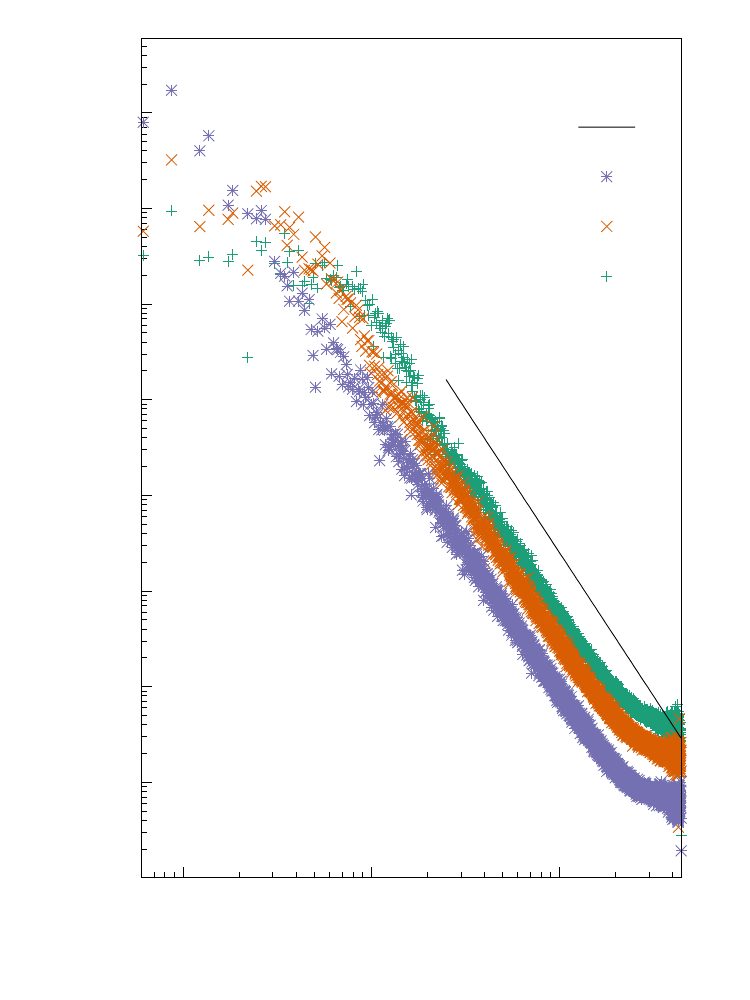}}%
    	\gplfronttext
    \end{picture}%
    \endgroup

	\caption{}
	\label{fig:modela:sf2d}
    \end{subfigure}%
    \begin{subfigure}{.5\textwidth}
    \centering
    
    \begingroup
    \inputencoding{cp1252}%
    \makeatletter
    \providecommand\color[2][]{%
    	\GenericError{(gnuplot) \space\space\space\@spaces}{%
    		Package color not loaded in conjunction with
    		terminal option `colourtext'%
    	}{See the gnuplot documentation for explanation.%
    	}{Either use 'blacktext' in gnuplot or load the package
    		color.sty in LaTeX.}%
    	\renewcommand\color[2][]{}%
    }%
    \providecommand\includegraphics[2][]{%
    	\GenericError{(gnuplot) \space\space\space\@spaces}{%
    		Package graphicx or graphics not loaded%
    	}{See the gnuplot documentation for explanation.%
    	}{The gnuplot epslatex terminal needs graphicx.sty or graphics.sty.}%
    	\renewcommand\includegraphics[2][]{}%
    }%
    \providecommand\rotatebox[2]{#2}%
    \@ifundefined{ifGPcolor}{%
    	\newif\ifGPcolor
    	\GPcolortrue
    }{}%
    \@ifundefined{ifGPblacktext}{%
    	\newif\ifGPblacktext
    	\GPblacktexttrue
    }{}%
    \let\gplgaddtomacro\g@addto@macro
    \gdef\gplbacktext{}%
    \gdef\gplfronttext{}%
    \makeatother
    \ifGPblacktext
    \def\colorrgb#1{}%
    \def\colorgray#1{}%
    \else
    \ifGPcolor
    \def\colorrgb#1{\color[rgb]{#1}}%
    \def\colorgray#1{\color[gray]{#1}}%
    \expandafter\def\csname LTw\endcsname{\color{white}}%
    \expandafter\def\csname LTb\endcsname{\color{black}}%
    \expandafter\def\csname LTa\endcsname{\color{black}}%
    \expandafter\def\csname LT0\endcsname{\color[rgb]{1,0,0}}%
    \expandafter\def\csname LT1\endcsname{\color[rgb]{0,1,0}}%
    \expandafter\def\csname LT2\endcsname{\color[rgb]{0,0,1}}%
    \expandafter\def\csname LT3\endcsname{\color[rgb]{1,0,1}}%
    \expandafter\def\csname LT4\endcsname{\color[rgb]{0,1,1}}%
    \expandafter\def\csname LT5\endcsname{\color[rgb]{1,1,0}}%
    \expandafter\def\csname LT6\endcsname{\color[rgb]{0,0,0}}%
    \expandafter\def\csname LT7\endcsname{\color[rgb]{1,0.3,0}}%
    \expandafter\def\csname LT8\endcsname{\color[rgb]{0.5,0.5,0.5}}%
    \else
    \def\colorrgb#1{\color{black}}%
    \def\colorgray#1{\color[gray]{#1}}%
    \expandafter\def\csname LTw\endcsname{\color{white}}%
    \expandafter\def\csname LTb\endcsname{\color{black}}%
    \expandafter\def\csname LTa\endcsname{\color{black}}%
    \expandafter\def\csname LT0\endcsname{\color{black}}%
    \expandafter\def\csname LT1\endcsname{\color{black}}%
    \expandafter\def\csname LT2\endcsname{\color{black}}%
    \expandafter\def\csname LT3\endcsname{\color{black}}%
    \expandafter\def\csname LT4\endcsname{\color{black}}%
    \expandafter\def\csname LT5\endcsname{\color{black}}%
    \expandafter\def\csname LT6\endcsname{\color{black}}%
    \expandafter\def\csname LT7\endcsname{\color{black}}%
    \expandafter\def\csname LT8\endcsname{\color{black}}%
    \fi
    \fi
    \setlength{\unitlength}{0.0500bp}%
    \ifx\gptboxheight\undefined%
    \newlength{\gptboxheight}%
    \newlength{\gptboxwidth}%
    \newsavebox{\gptboxtext}%
    \fi%
    \setlength{\fboxrule}{0.5pt}%
    \setlength{\fboxsep}{1pt}%
    \begin{picture}(4320.00,5760.00)%
    	\gplgaddtomacro\gplbacktext{%
    		\csname LTb\endcsname
    		\put(682,704){\makebox(0,0)[r]{\strut{}-4}}%
    		\put(682,1587){\makebox(0,0)[r]{\strut{}-2}}%
    		\put(682,2471){\makebox(0,0)[r]{\strut{}0}}%
    		\put(682,3354){\makebox(0,0)[r]{\strut{}2}}%
    		\put(682,4238){\makebox(0,0)[r]{\strut{}4}}%
    		\put(682,5121){\makebox(0,0)[r]{\strut{}6}}%
    		\put(1707,484){\makebox(0,0){\strut{}$0.1$}}%
    		\put(2984,484){\makebox(0,0){\strut{}$1$}}%
    	}%
    	\gplgaddtomacro\gplfronttext{%
    		\csname LTb\endcsname
    		\put(198,3121){\rotatebox{-270}{\makebox(0,0){\strut{}$S(k)$}}}%
    		\put(2368,154){\makebox(0,0){\strut{}$k$}}%
    		\put(2835,5333){\makebox(0,0){\strut{}Model A, 3D}}%
    		\csname LTb\endcsname
    		\put(3200,4170){\makebox(0,0)[r]{\strut{}50}}%
    		\csname LTb\endcsname
    		\put(3200,4456){\makebox(0,0)[r]{\strut{}200}}%
    		\csname LTb\endcsname
    		\put(3200,4742){\makebox(0,0)[r]{\strut{}2,000}}%
    		\csname LTb\endcsname
    		\put(3200,5028){\makebox(0,0)[r]{\strut{}$S(k) \sim k^{-4}$}}%
    	}%
    	\gplbacktext
    	\put(0,0){\includegraphics{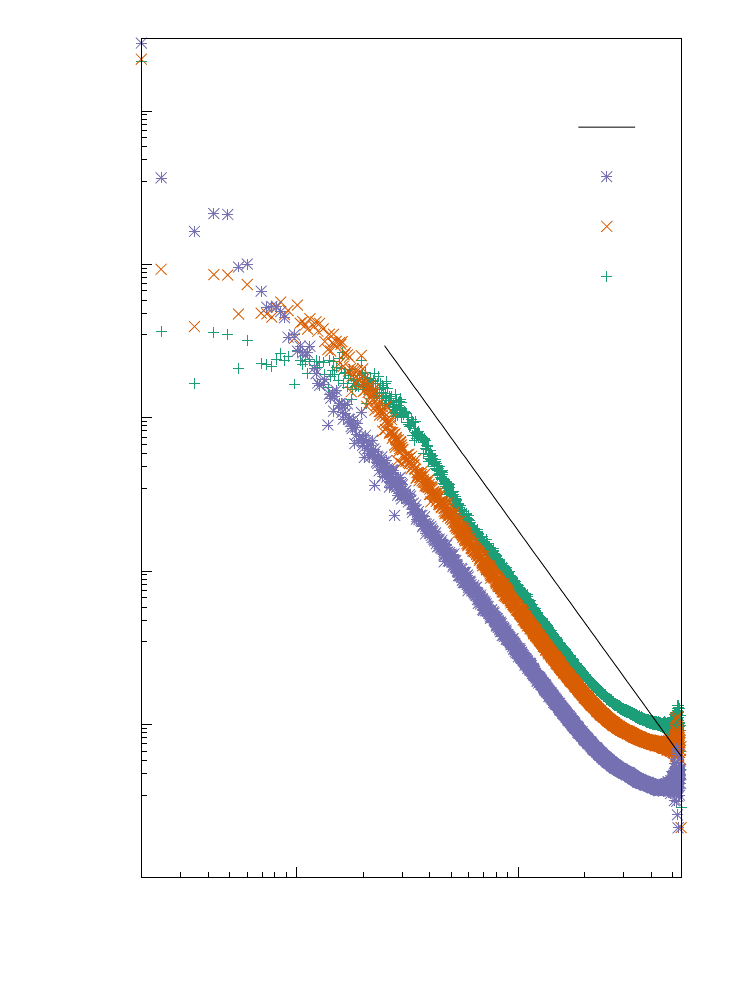}}%
    	\gplfronttext
    \end{picture}%
    \endgroup

	\caption{}
	\label{fig:modela:sf3d}
    \end{subfigure}
    \caption{The radially averaged structure factor, $S(\vec{k}) \! \! = \! | \hat{\psi}_{\vec{k}} |^2$, for the Allen--Cahn model \cite{Allen1975} in (\subref{fig:modela:sf2d}) 2D and (\subref{fig:modela:sf3d}) 3D, corresponding to the simulation parameters described in Figure~\ref{fig:modelab:a} and computed using the average of 10 simulations. The results demonstrate scaling to Porod's law~(solid line, Equation~(\ref{eq:porodslaw})) for $d=2$ and $d=3$, respectively~\cite{Puri1997}.}
    \label{fig:model-a_sf}
\end{figure}

\begin{figure}
    \centering
    \begin{subfigure}{.5\textwidth}
    \centering
    \begingroup
    \inputencoding{cp1252}%
    \makeatletter
    \providecommand\color[2][]{%
    	\GenericError{(gnuplot) \space\space\space\@spaces}{%
    		Package color not loaded in conjunction with
    		terminal option `colourtext'%
    	}{See the gnuplot documentation for explanation.%
    	}{Either use 'blacktext' in gnuplot or load the package
    		color.sty in LaTeX.}%
    	\renewcommand\color[2][]{}%
    }%
    \providecommand\includegraphics[2][]{%
    	\GenericError{(gnuplot) \space\space\space\@spaces}{%
    		Package graphicx or graphics not loaded%
    	}{See the gnuplot documentation for explanation.%
    	}{The gnuplot epslatex terminal needs graphicx.sty or graphics.sty.}%
    	\renewcommand\includegraphics[2][]{}%
    }%
    \providecommand\rotatebox[2]{#2}%
    \@ifundefined{ifGPcolor}{%
    	\newif\ifGPcolor
    	\GPcolortrue
    }{}%
    \@ifundefined{ifGPblacktext}{%
    	\newif\ifGPblacktext
    	\GPblacktexttrue
    }{}%
    \let\gplgaddtomacro\g@addto@macro
    \gdef\gplbacktext{}%
    \gdef\gplfronttext{}%
    \makeatother
    \ifGPblacktext
    \def\colorrgb#1{}%
    \def\colorgray#1{}%
    \else
    \ifGPcolor
    \def\colorrgb#1{\color[rgb]{#1}}%
    \def\colorgray#1{\color[gray]{#1}}%
    \expandafter\def\csname LTw\endcsname{\color{white}}%
    \expandafter\def\csname LTb\endcsname{\color{black}}%
    \expandafter\def\csname LTa\endcsname{\color{black}}%
    \expandafter\def\csname LT0\endcsname{\color[rgb]{1,0,0}}%
    \expandafter\def\csname LT1\endcsname{\color[rgb]{0,1,0}}%
    \expandafter\def\csname LT2\endcsname{\color[rgb]{0,0,1}}%
    \expandafter\def\csname LT3\endcsname{\color[rgb]{1,0,1}}%
    \expandafter\def\csname LT4\endcsname{\color[rgb]{0,1,1}}%
    \expandafter\def\csname LT5\endcsname{\color[rgb]{1,1,0}}%
    \expandafter\def\csname LT6\endcsname{\color[rgb]{0,0,0}}%
    \expandafter\def\csname LT7\endcsname{\color[rgb]{1,0.3,0}}%
    \expandafter\def\csname LT8\endcsname{\color[rgb]{0.5,0.5,0.5}}%
    \else
    \def\colorrgb#1{\color{black}}%
    \def\colorgray#1{\color[gray]{#1}}%
    \expandafter\def\csname LTw\endcsname{\color{white}}%
    \expandafter\def\csname LTb\endcsname{\color{black}}%
    \expandafter\def\csname LTa\endcsname{\color{black}}%
    \expandafter\def\csname LT0\endcsname{\color{black}}%
    \expandafter\def\csname LT1\endcsname{\color{black}}%
    \expandafter\def\csname LT2\endcsname{\color{black}}%
    \expandafter\def\csname LT3\endcsname{\color{black}}%
    \expandafter\def\csname LT4\endcsname{\color{black}}%
    \expandafter\def\csname LT5\endcsname{\color{black}}%
    \expandafter\def\csname LT6\endcsname{\color{black}}%
    \expandafter\def\csname LT7\endcsname{\color{black}}%
    \expandafter\def\csname LT8\endcsname{\color{black}}%
    \fi
    \fi
    \setlength{\unitlength}{0.0500bp}%
    \ifx\gptboxheight\undefined%
    \newlength{\gptboxheight}%
    \newlength{\gptboxwidth}%
    \newsavebox{\gptboxtext}%
    \fi%
    \setlength{\fboxrule}{0.5pt}%
    \setlength{\fboxsep}{1pt}%
    \begin{picture}(4320.00,5760.00)%
    	\gplgaddtomacro\gplbacktext{%
    		\csname LTb\endcsname
    		\put(682,704){\makebox(0,0)[r]{\strut{}-4}}%
    		\put(682,1671){\makebox(0,0)[r]{\strut{}-2}}%
    		\put(682,2638){\makebox(0,0)[r]{\strut{}0}}%
    		\put(682,3605){\makebox(0,0)[r]{\strut{}2}}%
    		\put(682,4572){\makebox(0,0)[r]{\strut{}4}}%
    		\put(682,5539){\makebox(0,0)[r]{\strut{}6}}%
    		\put(1054,484){\makebox(0,0){\strut{}$0.01$}}%
    		\put(2138,484){\makebox(0,0){\strut{}$0.1$}}%
    		\put(3221,484){\makebox(0,0){\strut{}$1$}}%
    	}%
    	\gplgaddtomacro\gplfronttext{%
    		\csname LTb\endcsname
    		\put(198,3121){\rotatebox{-270}{\makebox(0,0){\strut{}$S(k)$}}}%
    		\put(2368,154){\makebox(0,0){\strut{}$k$}}%
    		\put(2835,5333){\makebox(0,0){\strut{}Model B, 2D}}%
    		\csname LTb\endcsname
    		\put(3200,4170){\makebox(0,0)[r]{\strut{}5,000}}%
    		\csname LTb\endcsname
    		\put(3200,4456){\makebox(0,0)[r]{\strut{}20,000}}%
    		\csname LTb\endcsname
    		\put(3200,4742){\makebox(0,0)[r]{\strut{}200,000}}%
    		\csname LTb\endcsname
    		\put(3200,5028){\makebox(0,0)[r]{\strut{}$S(k) \sim k^{-3}$}}%
    	}%
    	\gplbacktext
    	\put(0,0){\includegraphics{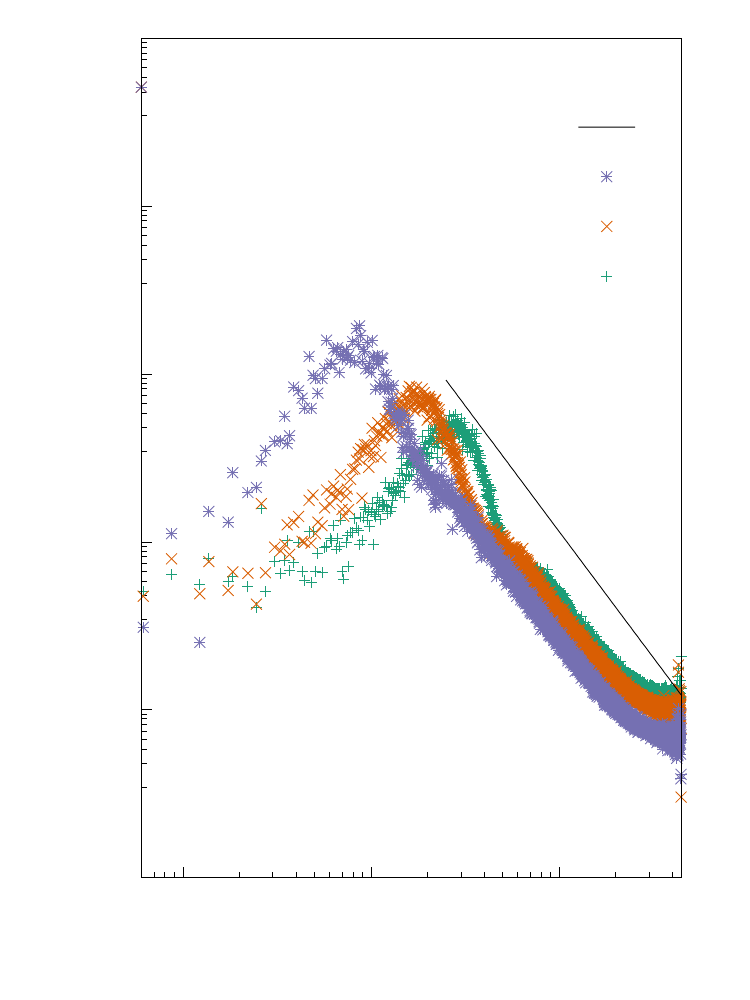}}%
    	\gplfronttext
    \end{picture}%
    \endgroup

	\caption{}
	\label{fig:modelb:sf2d}
    \end{subfigure}%
    \begin{subfigure}{.5\textwidth}
    \centering
    
    \begingroup
    \inputencoding{cp1252}%
    \makeatletter
    \providecommand\color[2][]{%
    	\GenericError{(gnuplot) \space\space\space\@spaces}{%
    		Package color not loaded in conjunction with
    		terminal option `colourtext'%
    	}{See the gnuplot documentation for explanation.%
    	}{Either use 'blacktext' in gnuplot or load the package
    		color.sty in LaTeX.}%
    	\renewcommand\color[2][]{}%
    }%
    \providecommand\includegraphics[2][]{%
    	\GenericError{(gnuplot) \space\space\space\@spaces}{%
    		Package graphicx or graphics not loaded%
    	}{See the gnuplot documentation for explanation.%
    	}{The gnuplot epslatex terminal needs graphicx.sty or graphics.sty.}%
    	\renewcommand\includegraphics[2][]{}%
    }%
    \providecommand\rotatebox[2]{#2}%
    \@ifundefined{ifGPcolor}{%
    	\newif\ifGPcolor
    	\GPcolortrue
    }{}%
    \@ifundefined{ifGPblacktext}{%
    	\newif\ifGPblacktext
    	\GPblacktexttrue
    }{}%
    \let\gplgaddtomacro\g@addto@macro
    \gdef\gplbacktext{}%
    \gdef\gplfronttext{}%
    \makeatother
    \ifGPblacktext
    \def\colorrgb#1{}%
    \def\colorgray#1{}%
    \else
    \ifGPcolor
    \def\colorrgb#1{\color[rgb]{#1}}%
    \def\colorgray#1{\color[gray]{#1}}%
    \expandafter\def\csname LTw\endcsname{\color{white}}%
    \expandafter\def\csname LTb\endcsname{\color{black}}%
    \expandafter\def\csname LTa\endcsname{\color{black}}%
    \expandafter\def\csname LT0\endcsname{\color[rgb]{1,0,0}}%
    \expandafter\def\csname LT1\endcsname{\color[rgb]{0,1,0}}%
    \expandafter\def\csname LT2\endcsname{\color[rgb]{0,0,1}}%
    \expandafter\def\csname LT3\endcsname{\color[rgb]{1,0,1}}%
    \expandafter\def\csname LT4\endcsname{\color[rgb]{0,1,1}}%
    \expandafter\def\csname LT5\endcsname{\color[rgb]{1,1,0}}%
    \expandafter\def\csname LT6\endcsname{\color[rgb]{0,0,0}}%
    \expandafter\def\csname LT7\endcsname{\color[rgb]{1,0.3,0}}%
    \expandafter\def\csname LT8\endcsname{\color[rgb]{0.5,0.5,0.5}}%
    \else
    \def\colorrgb#1{\color{black}}%
    \def\colorgray#1{\color[gray]{#1}}%
    \expandafter\def\csname LTw\endcsname{\color{white}}%
    \expandafter\def\csname LTb\endcsname{\color{black}}%
    \expandafter\def\csname LTa\endcsname{\color{black}}%
    \expandafter\def\csname LT0\endcsname{\color{black}}%
    \expandafter\def\csname LT1\endcsname{\color{black}}%
    \expandafter\def\csname LT2\endcsname{\color{black}}%
    \expandafter\def\csname LT3\endcsname{\color{black}}%
    \expandafter\def\csname LT4\endcsname{\color{black}}%
    \expandafter\def\csname LT5\endcsname{\color{black}}%
    \expandafter\def\csname LT6\endcsname{\color{black}}%
    \expandafter\def\csname LT7\endcsname{\color{black}}%
    \expandafter\def\csname LT8\endcsname{\color{black}}%
    \fi
    \fi
    \setlength{\unitlength}{0.0500bp}%
    \ifx\gptboxheight\undefined%
    \newlength{\gptboxheight}%
    \newlength{\gptboxwidth}%
    \newsavebox{\gptboxtext}%
    \fi%
    \setlength{\fboxrule}{0.5pt}%
    \setlength{\fboxsep}{1pt}%
    \begin{picture}(4320.00,5760.00)%
    	\gplgaddtomacro\gplbacktext{%
    		\csname LTb\endcsname
    		\put(682,823){\makebox(0,0)[r]{\strut{}-3}}%
    		\put(682,1360){\makebox(0,0)[r]{\strut{}-2}}%
    		\put(682,1897){\makebox(0,0)[r]{\strut{}-1}}%
    		\put(682,2435){\makebox(0,0)[r]{\strut{}0}}%
    		\put(682,2972){\makebox(0,0)[r]{\strut{}1}}%
    		\put(682,3509){\makebox(0,0)[r]{\strut{}2}}%
    		\put(682,4046){\makebox(0,0)[r]{\strut{}3}}%
    		\put(682,4583){\makebox(0,0)[r]{\strut{}4}}%
    		\put(682,5120){\makebox(0,0)[r]{\strut{}5}}%
    		\put(1394,484){\makebox(0,0){\strut{}$0.1$}}%
    		\put(2851,484){\makebox(0,0){\strut{}$1$}}%
    	}%
    	\gplgaddtomacro\gplfronttext{%
    		\csname LTb\endcsname
    		\put(198,3121){\rotatebox{-270}{\makebox(0,0){\strut{}$S(k)$}}}%
    		\put(2368,154){\makebox(0,0){\strut{}$k$}}%
    		\put(2835,5333){\makebox(0,0){\strut{}Model B, 3D}}%
    		\csname LTb\endcsname
    		\put(3200,4170){\makebox(0,0)[r]{\strut{}500}}%
    		\csname LTb\endcsname
    		\put(3200,4456){\makebox(0,0)[r]{\strut{}2,000}}%
    		\csname LTb\endcsname
    		\put(3200,4742){\makebox(0,0)[r]{\strut{}20,000}}%
    		\csname LTb\endcsname
    		\put(3200,5028){\makebox(0,0)[r]{\strut{}$S(k) \sim k^{-4}$}}%
    	}%
    	\gplbacktext
    	\put(0,0){\includegraphics{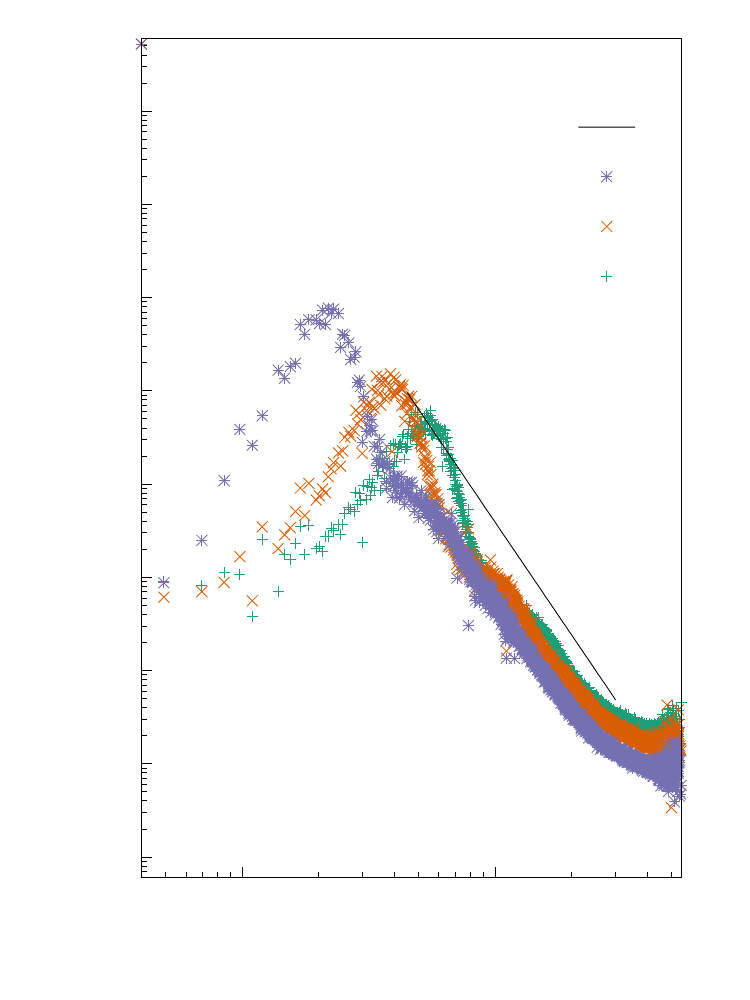}}%
    	\gplfronttext
    \end{picture}%
    \endgroup

	\caption{}
	\label{fig:modelb:sf3d}
    \end{subfigure}
    \caption{The radially averaged structure factor, $S(\vec{k}) \! = \! | \hat{\psi}_{\vec{k}} |^2$, for the Cahn--Hilliard model \cite{Cahn1958} in (\subref{fig:modelb:sf2d}) 2D and (\subref{fig:modelb:sf3d}) 3D, corresponding to the simulation parameters described in Figure~\ref{fig:modelab:b} and computed using the average of 10 simulations. The results demonstrate scaling to Porod's law~(solid line, Equation~(\ref{eq:porodslaw})) for $d=2$ and $d=3$, respectively~\cite{Puri1997}. }
    \label{fig:model-b_sf}
\end{figure}

\begin{figure}
\begin{minipage}{0.48\textwidth}
    \centering
    \begin{subfigure}{\textwidth}
    \centering
    \begingroup
    \inputencoding{cp1252}%
    \makeatletter
    \providecommand\color[2][]{%
    	\GenericError{(gnuplot) \space\space\space\@spaces}{%
    		Package color not loaded in conjunction with
    		terminal option `colourtext'%
    	}{See the gnuplot documentation for explanation.%
    	}{Either use 'blacktext' in gnuplot or load the package
    		color.sty in LaTeX.}%
    	\renewcommand\color[2][]{}%
    }%
    \providecommand\includegraphics[2][]{%
    	\GenericError{(gnuplot) \space\space\space\@spaces}{%
    		Package graphicx or graphics not loaded%
    	}{See the gnuplot documentation for explanation.%
    	}{The gnuplot epslatex terminal needs graphicx.sty or graphics.sty.}%
    	\renewcommand\includegraphics[2][]{}%
    }%
    \providecommand\rotatebox[2]{#2}%
    \@ifundefined{ifGPcolor}{%
    	\newif\ifGPcolor
    	\GPcolortrue
    }{}%
    \@ifundefined{ifGPblacktext}{%
    	\newif\ifGPblacktext
    	\GPblacktexttrue
    }{}%
    \let\gplgaddtomacro\g@addto@macro
    \gdef\gplbacktext{}%
    \gdef\gplfronttext{}%
    \makeatother
    \ifGPblacktext
    \def\colorrgb#1{}%
    \def\colorgray#1{}%
    \else
    \ifGPcolor
    \def\colorrgb#1{\color[rgb]{#1}}%
    \def\colorgray#1{\color[gray]{#1}}%
    \expandafter\def\csname LTw\endcsname{\color{white}}%
    \expandafter\def\csname LTb\endcsname{\color{black}}%
    \expandafter\def\csname LTa\endcsname{\color{black}}%
    \expandafter\def\csname LT0\endcsname{\color[rgb]{1,0,0}}%
    \expandafter\def\csname LT1\endcsname{\color[rgb]{0,1,0}}%
    \expandafter\def\csname LT2\endcsname{\color[rgb]{0,0,1}}%
    \expandafter\def\csname LT3\endcsname{\color[rgb]{1,0,1}}%
    \expandafter\def\csname LT4\endcsname{\color[rgb]{0,1,1}}%
    \expandafter\def\csname LT5\endcsname{\color[rgb]{1,1,0}}%
    \expandafter\def\csname LT6\endcsname{\color[rgb]{0,0,0}}%
    \expandafter\def\csname LT7\endcsname{\color[rgb]{1,0.3,0}}%
    \expandafter\def\csname LT8\endcsname{\color[rgb]{0.5,0.5,0.5}}%
    \else
    \def\colorrgb#1{\color{black}}%
    \def\colorgray#1{\color[gray]{#1}}%
    \expandafter\def\csname LTw\endcsname{\color{white}}%
    \expandafter\def\csname LTb\endcsname{\color{black}}%
    \expandafter\def\csname LTa\endcsname{\color{black}}%
    \expandafter\def\csname LT0\endcsname{\color{black}}%
    \expandafter\def\csname LT1\endcsname{\color{black}}%
    \expandafter\def\csname LT2\endcsname{\color{black}}%
    \expandafter\def\csname LT3\endcsname{\color{black}}%
    \expandafter\def\csname LT4\endcsname{\color{black}}%
    \expandafter\def\csname LT5\endcsname{\color{black}}%
    \expandafter\def\csname LT6\endcsname{\color{black}}%
    \expandafter\def\csname LT7\endcsname{\color{black}}%
    \expandafter\def\csname LT8\endcsname{\color{black}}%
    \fi
    \fi
    \setlength{\unitlength}{0.0500bp}%
    \ifx\gptboxheight\undefined%
    \newlength{\gptboxheight}%
    \newlength{\gptboxwidth}%
    \newsavebox{\gptboxtext}%
    \fi%
    \setlength{\fboxrule}{0.5pt}%
    \setlength{\fboxsep}{1pt}%
    \begin{picture}(5182.00,1872.00)%
    	\gplgaddtomacro\gplbacktext{%
    	}%
    	\gplgaddtomacro\gplfronttext{%
    		\csname LTb\endcsname
    		\put(1714,1924){\makebox(0,0)[r]{\strut{}$25,000$}}%
    	}%
    	\gplgaddtomacro\gplbacktext{%
    	}%
    	\gplgaddtomacro\gplfronttext{%
    		\csname LTb\endcsname
    		\put(3341,1924){\makebox(0,0)[r]{\strut{}$100,000$}}%
    	}%
    	\gplgaddtomacro\gplbacktext{%
    	}%
    	\gplgaddtomacro\gplfronttext{%
    		\csname LTb\endcsname
    		\put(4963,1924){\makebox(0,0)[r]{\strut{}$1,000,000$}}%
    		\csname LTb\endcsname
    		\put(218,0){\makebox(0,0){\strut{}$-1$}}%
    		\put(1404,0){\makebox(0,0){\strut{}$-0.5$}}%
    		\put(2590,0){\makebox(0,0){\strut{}$0$}}%
    		\put(3776,0){\makebox(0,0){\strut{}$0.5$}}%
    		\put(4963,0){\makebox(0,0){\strut{}$1$}}%
    	}%
    	\gplbacktext
    	\put(0,0){\includegraphics{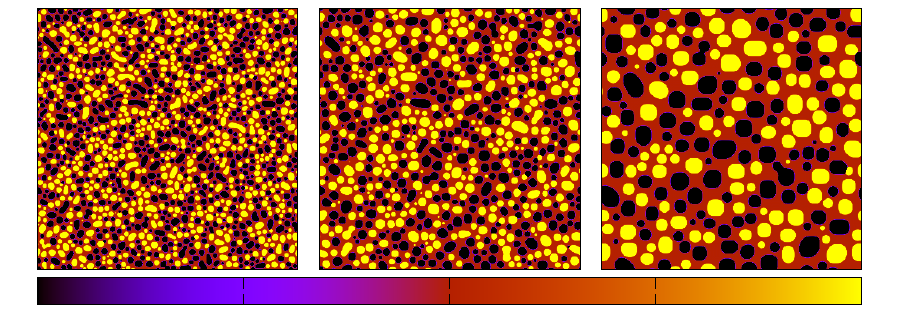}}%
    	\gplfronttext
    \end{picture}%
    \endgroup
    \vspace{15pt}
    \begingroup
    \inputencoding{cp1252}%
    \makeatletter
    \providecommand\color[2][]{%
    	\GenericError{(gnuplot) \space\space\space\@spaces}{%
    		Package color not loaded in conjunction with
    		terminal option `colourtext'%
    	}{See the gnuplot documentation for explanation.%
    	}{Either use 'blacktext' in gnuplot or load the package
    		color.sty in LaTeX.}%
    	\renewcommand\color[2][]{}%
    }%
    \providecommand\includegraphics[2][]{%
    	\GenericError{(gnuplot) \space\space\space\@spaces}{%
    		Package graphicx or graphics not loaded%
    	}{See the gnuplot documentation for explanation.%
    	}{The gnuplot epslatex terminal needs graphicx.sty or graphics.sty.}%
    	\renewcommand\includegraphics[2][]{}%
    }%
    \providecommand\rotatebox[2]{#2}%
    \@ifundefined{ifGPcolor}{%
    	\newif\ifGPcolor
    	\GPcolortrue
    }{}%
    \@ifundefined{ifGPblacktext}{%
    	\newif\ifGPblacktext
    	\GPblacktexttrue
    }{}%
    \let\gplgaddtomacro\g@addto@macro
    \gdef\gplbacktext{}%
    \gdef\gplfronttext{}%
    \makeatother
    \ifGPblacktext
    \def\colorrgb#1{}%
    \def\colorgray#1{}%
    \else
    \ifGPcolor
    \def\colorrgb#1{\color[rgb]{#1}}%
    \def\colorgray#1{\color[gray]{#1}}%
    \expandafter\def\csname LTw\endcsname{\color{white}}%
    \expandafter\def\csname LTb\endcsname{\color{black}}%
    \expandafter\def\csname LTa\endcsname{\color{black}}%
    \expandafter\def\csname LT0\endcsname{\color[rgb]{1,0,0}}%
    \expandafter\def\csname LT1\endcsname{\color[rgb]{0,1,0}}%
    \expandafter\def\csname LT2\endcsname{\color[rgb]{0,0,1}}%
    \expandafter\def\csname LT3\endcsname{\color[rgb]{1,0,1}}%
    \expandafter\def\csname LT4\endcsname{\color[rgb]{0,1,1}}%
    \expandafter\def\csname LT5\endcsname{\color[rgb]{1,1,0}}%
    \expandafter\def\csname LT6\endcsname{\color[rgb]{0,0,0}}%
    \expandafter\def\csname LT7\endcsname{\color[rgb]{1,0.3,0}}%
    \expandafter\def\csname LT8\endcsname{\color[rgb]{0.5,0.5,0.5}}%
    \else
    \def\colorrgb#1{\color{black}}%
    \def\colorgray#1{\color[gray]{#1}}%
    \expandafter\def\csname LTw\endcsname{\color{white}}%
    \expandafter\def\csname LTb\endcsname{\color{black}}%
    \expandafter\def\csname LTa\endcsname{\color{black}}%
    \expandafter\def\csname LT0\endcsname{\color{black}}%
    \expandafter\def\csname LT1\endcsname{\color{black}}%
    \expandafter\def\csname LT2\endcsname{\color{black}}%
    \expandafter\def\csname LT3\endcsname{\color{black}}%
    \expandafter\def\csname LT4\endcsname{\color{black}}%
    \expandafter\def\csname LT5\endcsname{\color{black}}%
    \expandafter\def\csname LT6\endcsname{\color{black}}%
    \expandafter\def\csname LT7\endcsname{\color{black}}%
    \expandafter\def\csname LT8\endcsname{\color{black}}%
    \fi
    \fi
    \setlength{\unitlength}{0.0500bp}%
    \ifx\gptboxheight\undefined%
    \newlength{\gptboxheight}%
    \newlength{\gptboxwidth}%
    \newsavebox{\gptboxtext}%
    \fi%
    \setlength{\fboxrule}{0.5pt}%
    \setlength{\fboxsep}{1pt}%
    \begin{picture}(5182.00,1872.00)%
    	\gplgaddtomacro\gplbacktext{%
    	}%
    	\gplgaddtomacro\gplfronttext{%
    		\csname LTb\endcsname
    		\put(1714,1924){\makebox(0,0)[r]{\strut{}$25,000$}}%
    	}%
    	\gplgaddtomacro\gplbacktext{%
    	}%
    	\gplgaddtomacro\gplfronttext{%
    		\csname LTb\endcsname
    		\put(3341,1924){\makebox(0,0)[r]{\strut{}$100,000$}}%
    	}%
    	\gplgaddtomacro\gplbacktext{%
    	}%
    	\gplgaddtomacro\gplfronttext{%
    		\csname LTb\endcsname
    		\put(4963,1924){\makebox(0,0)[r]{\strut{}$1,000,000$}}%
    		\csname LTb\endcsname
    		\put(218,0){\makebox(0,0){\strut{}$-1$}}%
    		\put(1404,0){\makebox(0,0){\strut{}$-0.5$}}%
    		\put(2590,0){\makebox(0,0){\strut{}$0$}}%
    		\put(3776,0){\makebox(0,0){\strut{}$0.5$}}%
    		\put(4963,0){\makebox(0,0){\strut{}$1$}}%
    	}%
    	\gplbacktext
    	\put(0,0){\includegraphics{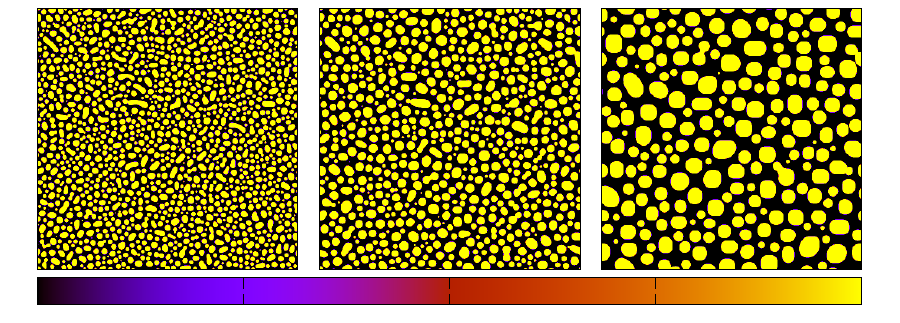}}%
    	\gplfronttext
    \end{picture}%
    \endgroup
    
    \caption{}
    \label{fig:modelc:2d}
    \end{subfigure}
    \begin{subfigure}{\textwidth}
    \centering
    \includegraphics[width=0.33\textwidth]{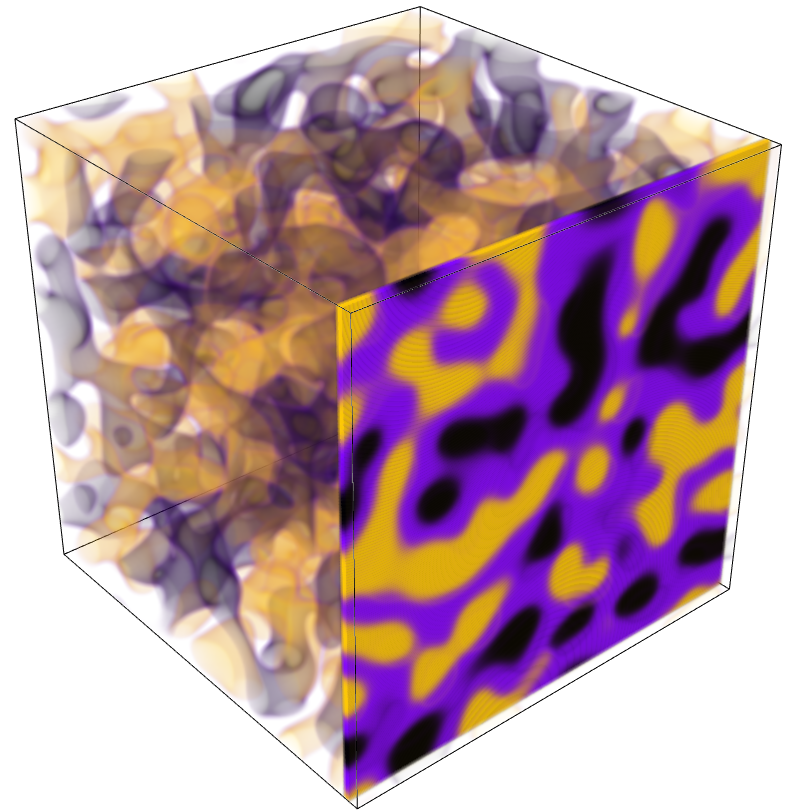}\hfill%
    \includegraphics[width=0.33\textwidth]{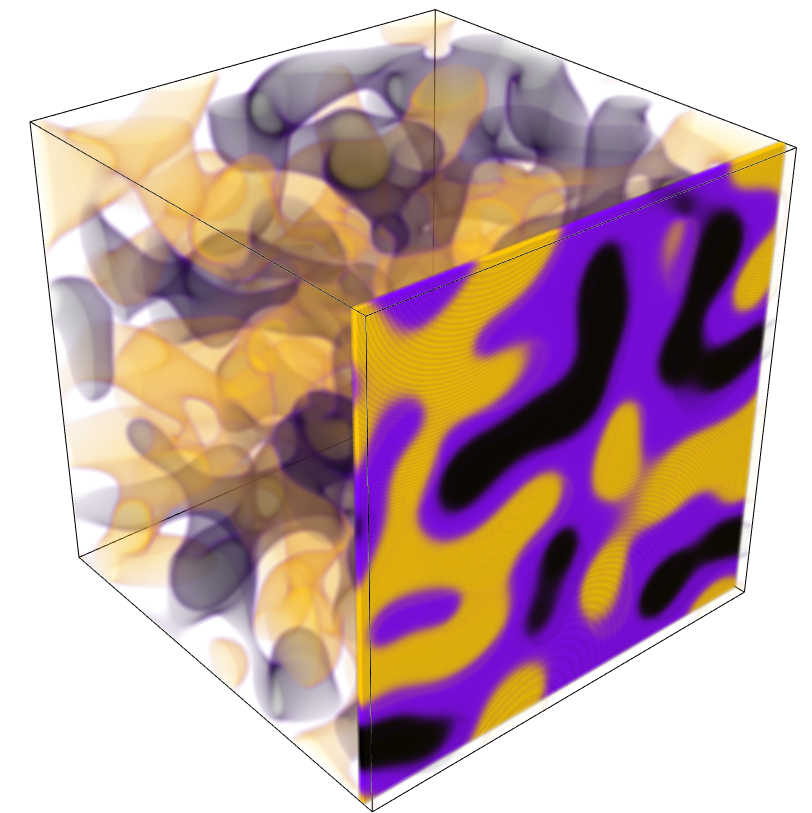}\hfill%
    \includegraphics[width=0.33\textwidth]{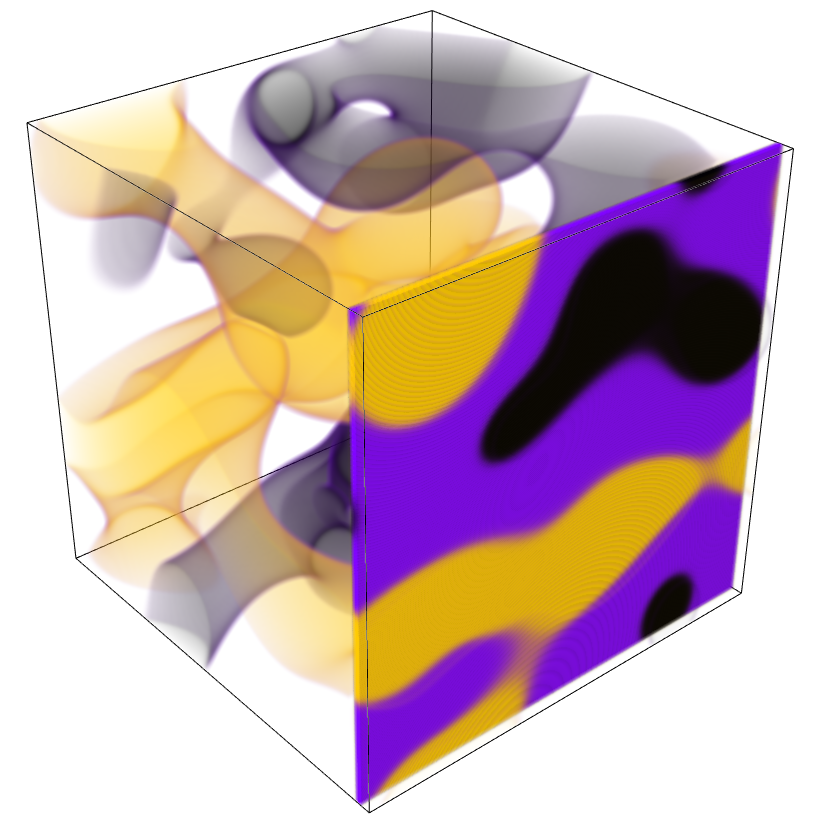}
    \includegraphics[width=0.33\textwidth]{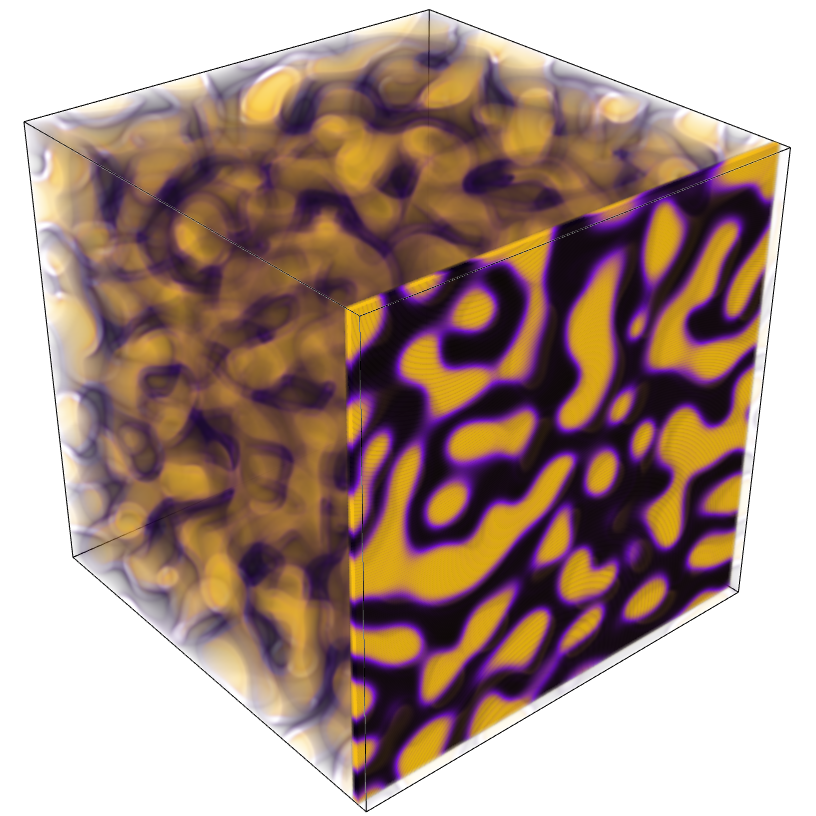}\hfill%
    \includegraphics[width=0.33\textwidth]{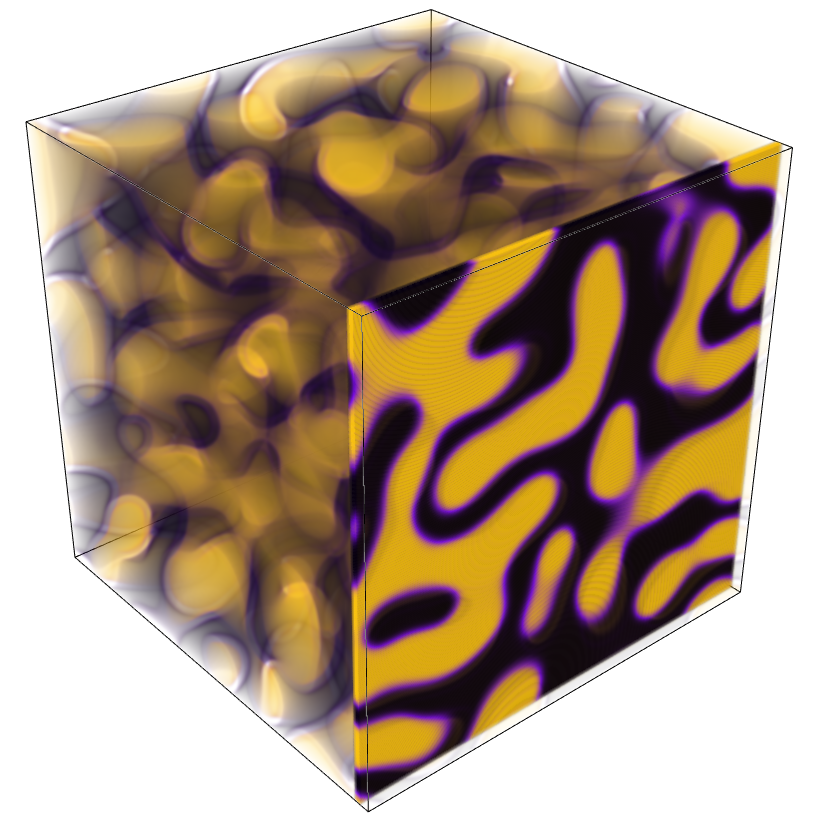}\hfill%
    \includegraphics[width=0.33\textwidth]{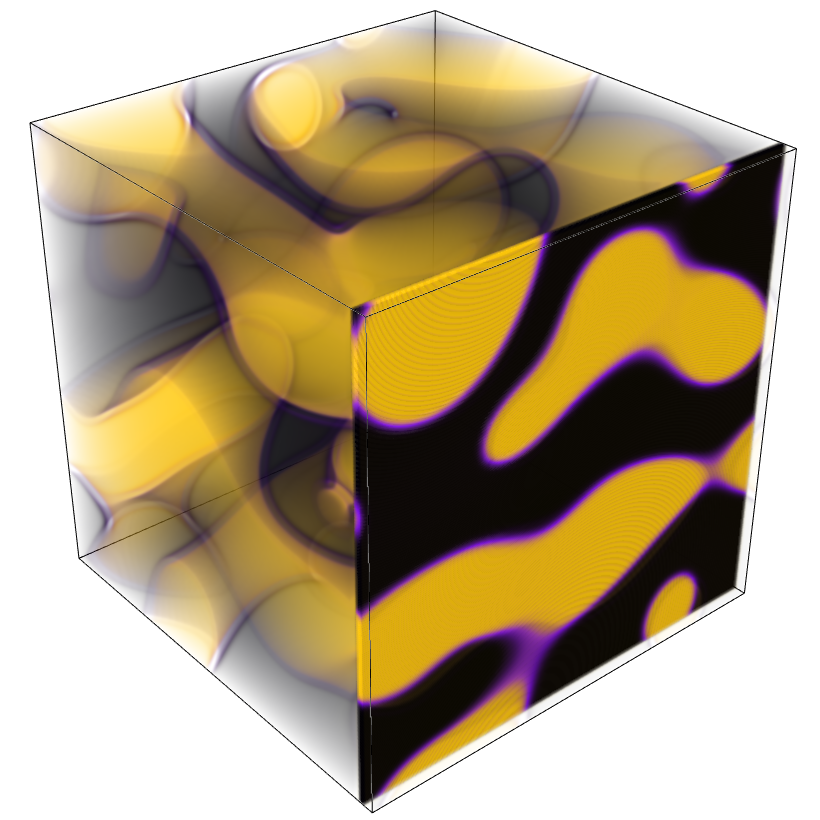}
    \caption{}
    \label{fig:modelc:3d}
    \end{subfigure}
    \caption{Snapshots from simulations of 2D ($1024 \! \times \!1024$) and 3D ($128 \! \times \! 128 \! \times \! 128$) Model C \cite{Elder1994}, Equation~(\ref{eq:modelc}):
    (\subref{fig:modelc:2d}) the 2D system is shown at three intervals at solution index 25,000, 100,000 and 1,000,000.  (\subref{fig:modelc:3d}) the 3D system  visualized using VTK \cite{vtk}, with a cross-section highlighted for visibility, at three intervals at solution index 2,500, 10,000 and 100,000.
    The simulations use a time step of $\Delta t = 0.025$, and were initially seeded with random values between -1 and 1. 
    The first row illustrates the evolution of the non-conserved field coalescing into droplet formations.
The conserved field illustrated on the second row represents the density of the material.
    }
    \label{fig:modelc}
\end{minipage}\hfill%
\begin{minipage}{0.48\textwidth}
    \centering
    \begin{subfigure}{\textwidth}
    \centering
    \begingroup
    \inputencoding{cp1252}%
    \makeatletter
    \providecommand\color[2][]{%
    	\GenericError{(gnuplot) \space\space\space\@spaces}{%
    		Package color not loaded in conjunction with
    		terminal option `colourtext'%
    	}{See the gnuplot documentation for explanation.%
    	}{Either use 'blacktext' in gnuplot or load the package
    		color.sty in LaTeX.}%
    	\renewcommand\color[2][]{}%
    }%
    \providecommand\includegraphics[2][]{%
    	\GenericError{(gnuplot) \space\space\space\@spaces}{%
    		Package graphicx or graphics not loaded%
    	}{See the gnuplot documentation for explanation.%
    	}{The gnuplot epslatex terminal needs graphicx.sty or graphics.sty.}%
    	\renewcommand\includegraphics[2][]{}%
    }%
    \providecommand\rotatebox[2]{#2}%
    \@ifundefined{ifGPcolor}{%
    	\newif\ifGPcolor
    	\GPcolortrue
    }{}%
    \@ifundefined{ifGPblacktext}{%
    	\newif\ifGPblacktext
    	\GPblacktexttrue
    }{}%
    \let\gplgaddtomacro\g@addto@macro
    \gdef\gplbacktext{}%
    \gdef\gplfronttext{}%
    \makeatother
    \ifGPblacktext
    \def\colorrgb#1{}%
    \def\colorgray#1{}%
    \else
    \ifGPcolor
    \def\colorrgb#1{\color[rgb]{#1}}%
    \def\colorgray#1{\color[gray]{#1}}%
    \expandafter\def\csname LTw\endcsname{\color{white}}%
    \expandafter\def\csname LTb\endcsname{\color{black}}%
    \expandafter\def\csname LTa\endcsname{\color{black}}%
    \expandafter\def\csname LT0\endcsname{\color[rgb]{1,0,0}}%
    \expandafter\def\csname LT1\endcsname{\color[rgb]{0,1,0}}%
    \expandafter\def\csname LT2\endcsname{\color[rgb]{0,0,1}}%
    \expandafter\def\csname LT3\endcsname{\color[rgb]{1,0,1}}%
    \expandafter\def\csname LT4\endcsname{\color[rgb]{0,1,1}}%
    \expandafter\def\csname LT5\endcsname{\color[rgb]{1,1,0}}%
    \expandafter\def\csname LT6\endcsname{\color[rgb]{0,0,0}}%
    \expandafter\def\csname LT7\endcsname{\color[rgb]{1,0.3,0}}%
    \expandafter\def\csname LT8\endcsname{\color[rgb]{0.5,0.5,0.5}}%
    \else
    \def\colorrgb#1{\color{black}}%
    \def\colorgray#1{\color[gray]{#1}}%
    \expandafter\def\csname LTw\endcsname{\color{white}}%
    \expandafter\def\csname LTb\endcsname{\color{black}}%
    \expandafter\def\csname LTa\endcsname{\color{black}}%
    \expandafter\def\csname LT0\endcsname{\color{black}}%
    \expandafter\def\csname LT1\endcsname{\color{black}}%
    \expandafter\def\csname LT2\endcsname{\color{black}}%
    \expandafter\def\csname LT3\endcsname{\color{black}}%
    \expandafter\def\csname LT4\endcsname{\color{black}}%
    \expandafter\def\csname LT5\endcsname{\color{black}}%
    \expandafter\def\csname LT6\endcsname{\color{black}}%
    \expandafter\def\csname LT7\endcsname{\color{black}}%
    \expandafter\def\csname LT8\endcsname{\color{black}}%
    \fi
    \fi
    \setlength{\unitlength}{0.0500bp}%
    \ifx\gptboxheight\undefined%
    \newlength{\gptboxheight}%
    \newlength{\gptboxwidth}%
    \newsavebox{\gptboxtext}%
    \fi%
    \setlength{\fboxrule}{0.5pt}%
    \setlength{\fboxsep}{1pt}%
    \begin{picture}(5182.00,4896.00)%
    	\gplgaddtomacro\gplbacktext{%
    		\csname LTb\endcsname
    		\put(671,979){\makebox(0,0)[r]{\strut{}$0$}}%
    		\put(671,1400){\makebox(0,0)[r]{\strut{}$100$}}%
    		\put(671,1822){\makebox(0,0)[r]{\strut{}$200$}}%
    		\put(671,2243){\makebox(0,0)[r]{\strut{}$300$}}%
    		\put(671,2664){\makebox(0,0)[r]{\strut{}$400$}}%
    		\put(671,3086){\makebox(0,0)[r]{\strut{}$500$}}%
    		\put(671,3507){\makebox(0,0)[r]{\strut{}$600$}}%
    		\put(671,3928){\makebox(0,0)[r]{\strut{}$700$}}%
    		\put(671,4350){\makebox(0,0)[r]{\strut{}$800$}}%
    		\put(803,759){\makebox(0,0){\strut{}$0$}}%
    		\put(1248,759){\makebox(0,0){\strut{}$100$}}%
    		\put(1694,759){\makebox(0,0){\strut{}$200$}}%
    		\put(2139,759){\makebox(0,0){\strut{}$300$}}%
    		\put(2584,759){\makebox(0,0){\strut{}$400$}}%
    		\put(3029,759){\makebox(0,0){\strut{}$500$}}%
    		\put(3475,759){\makebox(0,0){\strut{}$600$}}%
    		\put(3920,759){\makebox(0,0){\strut{}$700$}}%
    		\put(4365,759){\makebox(0,0){\strut{}$800$}}%
    	}%
    	\gplgaddtomacro\gplfronttext{%
    		\csname LTb\endcsname
    		\put(803,181){\makebox(0,0){\strut{}$-1$}}%
    		\put(1696,181){\makebox(0,0){\strut{}$-0.5$}}%
    		\put(2590,181){\makebox(0,0){\strut{}$0$}}%
    		\put(3483,181){\makebox(0,0){\strut{}$0.5$}}%
    		\put(4377,181){\makebox(0,0){\strut{}$1$}}%
    	}%
    	\gplbacktext
    	\put(0,0){\includegraphics{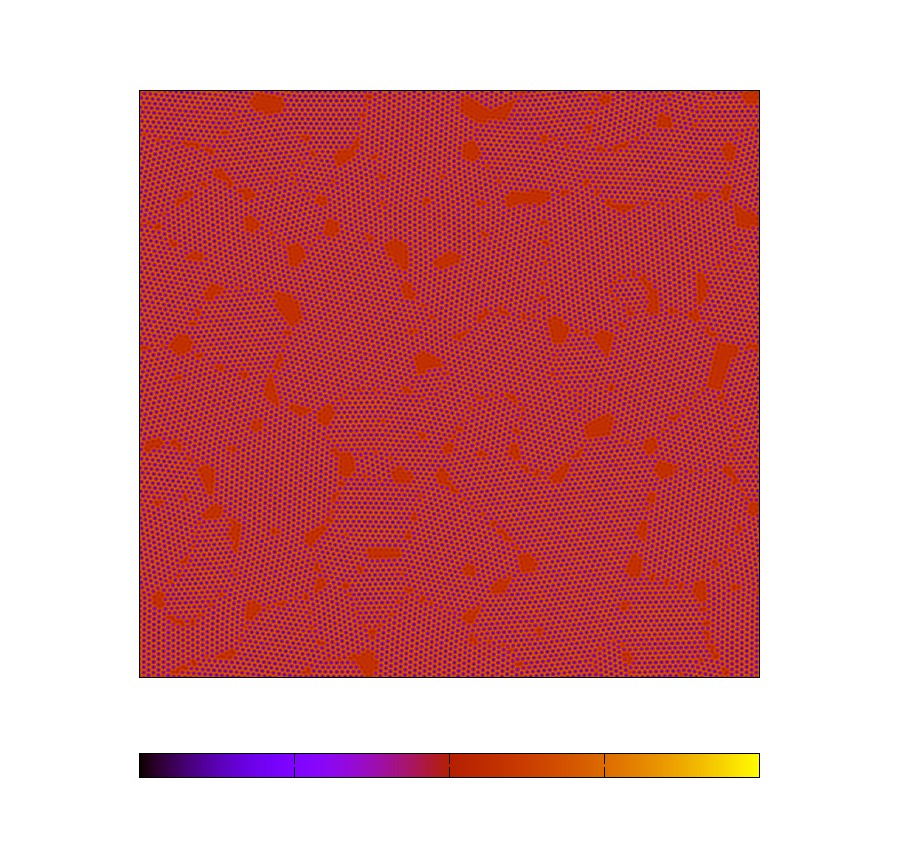}}%
    	\gplfronttext
    \end{picture}%
    \endgroup
    
    \caption{}
    \label{fig:modelpfc:2d}
    \end{subfigure}
    \begin{subfigure}{\textwidth}
    \centering
    \includegraphics[width=\textwidth]{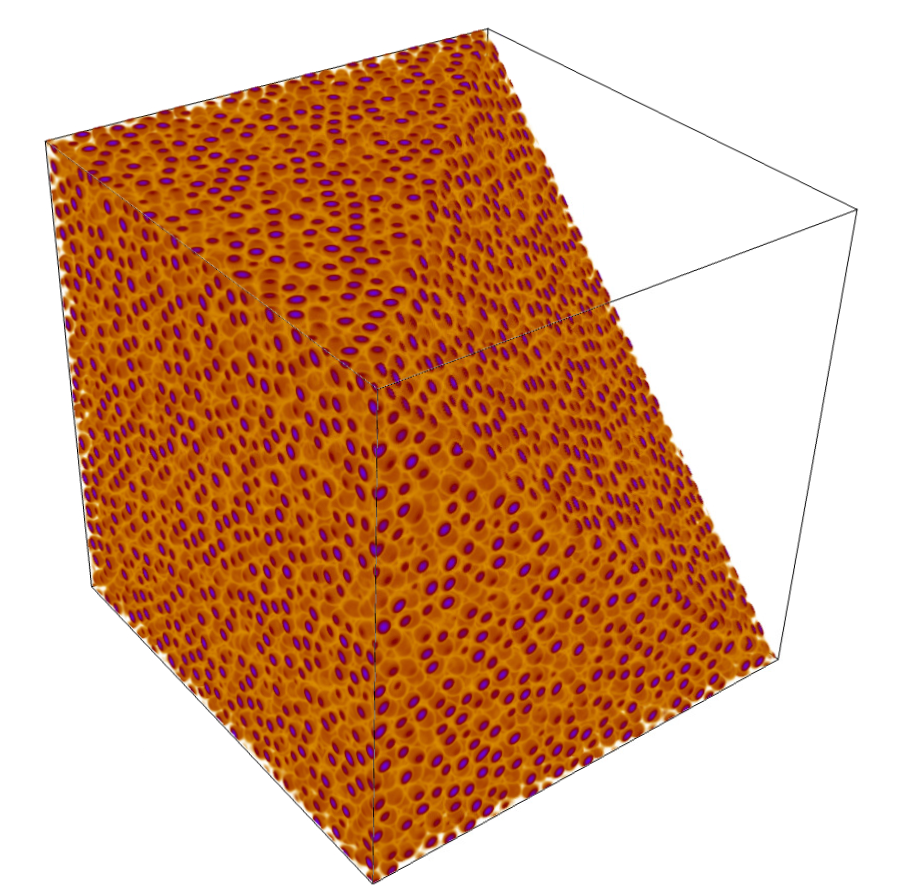}
    \caption{}
    \label{fig:modelpfc:3d}
    \end{subfigure}
    \caption{Snapshots from simulations of the phase-field crystal model. Parameters were taken from from Elder et al.~\cite{Elder2002}, $h= \pi/4$ and $\Delta t = 0.01$, and the constants in the dynamical equation (Equation~(\ref{eq:pfc})) were chosen to be $q_0=1$ and $\epsilon=0.1$. (\subref{fig:modelpfc:2d}) Shows the 2D ($1024\times1024$) simulation at 700,000 iterations, representing approximately 700 diffusion time-lengths. (\subref{fig:modelpfc:3d}) shows the visualization of the 3D system after 70,000 iterations, where a view of the interior is provided by removing a portion of the system along a sloped plane.
    }
    \label{fig:pfc}
\end{minipage}
\end{figure}

The initial conditions of models A, B and C are uniformly distributed noise, and the equation parameters $c_1$, $c_2$, $c_3$, $c_4$ and $c_5$ are set to unity. For the phase-field crystal model, 128 randomly arranged seeds containing large fluctuations are initially distributed throughout the system, and the parameters of the equation and simulation were selected from Elder et al.~\cite{Elder2002}.
The simulation results for the non-conserved Allen--Cahn model (Model A, Equation~(\ref{eq:modela})) are displayed in Figure~\ref{fig:modelab:a} and results for the conserved Cahn--Hilliard model (Model B, Equation~(\ref{eq:modela})) are displayed in Figure~\ref{fig:modelab:b}. The structure factor results of these two models are presented in Figure~\ref{fig:model-a_sf} and Figure~\ref{fig:model-b_sf}, respectively. The results for the eutectic model consisting of two coupled equations of motion~\cite{Elder1994} (Model C, Equation~(\ref{eq:modelc})) are shown in Figure~\ref{fig:modelc} and the phase-field crystal model~\cite{Elder2002} (Equation~(\ref{eq:pfc})) with a conserved field is displayed in Figure~\ref{fig:pfc}. 

\begin{figure}
    \centering
\begin{subfigure}{1.\textwidth}
    \footnotesize
    \begin{lstlisting}
MODEL(MA, (SCALAR),
  MODEL_DEF(
    dpsi = lap(psi) + (c1 - c2 * psi * psi) * psi))
    \end{lstlisting}
    \caption{}
    \label{fig:modeldef:a}
\end{subfigure}\hfill%
\begin{subfigure}{1.0\textwidth}
    \footnotesize
    \begin{lstlisting}
MODEL(MB, (SCALAR),
  MODEL_DEF(
    dpsi = -bilap(m) - lap((c1 - c2 * m * m) * m)))
    \end{lstlisting}
    \caption{}
    \label{fig:modeldef:b}
\end{subfigure}
\begin{subfigure}{1.0\textwidth}
    \footnotesize
    \begin{lstlisting}
MODEL(MC, (SCALAR, SCALAR),
  MODEL_PREAMBLE_DEF(
    ( auto psi3 = c2 * psi * psi * psi;
      auto m3 = c4 * m * m * m; ),
    dpsi = lap(psi) + c1 * psi - psi3 + lit(2.) * c5 * psi * m,
    dm = -bilap(m) - lap(c3 * m - m3 + c5 * psi * psi)))
    \end{lstlisting}
    \caption{}
    \label{fig:modeldef:c}
\end{subfigure}\hfill%
\begin{subfigure}{1.0\textwidth}
\footnotesize
\begin{lstlisting}
PFC_TYPE(PC, 
  DEFAULTS(
    DEFAULT_DYNAMIC(PFC_CONSERVED)
  ), 
  (SCALAR))
   
\end{lstlisting}
    \caption{}
    \label{fig:modeldef:pfc}
\end{subfigure}
    \caption{Macro implementations of (\subref{fig:modeldef:a}) the Allen--Cahn model \cite{Allen1975} from Equation~(\ref{eq:modela}),  (\subref{fig:modeldef:b}) the Cahn--Hilliard model \cite{Cahn1958} from Equation~(\ref{eq:modelb}),  and  (\subref{fig:modeldef:c}) Model C \cite{Elder1994} from Equation~(\ref{eq:modelc}). 
    The order parameter names in the macro specification are chosen to correspond to the variable names in the respective equations of motion, and the keys \lstinline{c1} to \lstinline{c5} correspond to the coefficients $c_1$ to $c_5$. (d) The phase-field crystal model (Equation~\ref{eq:pfc}) specification uses different macro keywords that allow selecting the phase-field crystal model as a type of problem, dispensing with equation specification and allowing selection of the dynamics.
    }
    
    \label{fig:models_abc_definitions}
\end{figure}%

\subsection{Performance}

Performance was measured on three different hardware and operating system platforms using Models A and C with
the solution taken after 10,000 iterations. The performance was measured by the runtime length of the entire program execution,
in seconds. Since this includes all program activity between program initialization and termination, the data includes both time spent generating the spectral form and printing the final results to a file. All recorded measurements are listed in Table~\ref{table:performance}, and the details of the hardware platforms are listed in Table~\ref{table:hardware}.

Data was collected by repeating the simulations 10 times.
From the system size, it is expected that the second test case is approximately 16 times longer than the first, and the third test case is similar to the second test case, for each model. Additionally, it is expected that Model C results are approximately twice as long as Model A (although the derivatives are of a higher order, and there is a coupling term). The data shows that typically the second test is much longer than the first test, which is the result of the ability for the processor to perform caching on the smaller system. Otherwise, these results correspond well to expectations.

\begin{table}
    \caption{Runtime data was recorded for three different hardware and operating system environments, for two models and three different grid sizes. The entries of the table show the results of taking the minimum and maximum runtime values (in seconds) for each individual configuration across 10 individual simulations, in the format ``minimum value/maximum value''. The hardware and operating system specifications of the environments used are given in Table~\ref{table:hardware}.}
    \footnotesize
    \centering
    \begin{tabular}{|c|l|l|l|l|l|l|}
        \hline
        \multirow{2}{*}{Label} & \multicolumn{3}{c}{Model A} & \multicolumn{3}{|c|}{Model C} \\
        & {$128 \times 128$} & {$512 \times 512$} & {$64 \times 64 \times 64$} & {$128 \times 128$} & {$512 \times 512$} & {$64 \times 64 \times 64$} \\ \hline
        Win7 & 2.0/2.4 & 33.9/36.4 & 37.1/40.1 & 5.0/5.7 & 118.5/125.1 & 113.9/122.6 \\ 
        Win10 & 1.7/1.8 & 32.6/33.4 & 31.1/31.6 & 4.1/4.2 & 102.8/106.3 & 96.7/97.3 \\
        Arch & 2.0/2.0 & 49.8/50.1 & 45.8/46.2 & 5.5/5.7 & 175.0/177.8 & 158.7/160.3 \\
        \hline
    \end{tabular}
    
    \label{table:performance}
\end{table}

\begin{table}
    \caption{The hardware and operating system specifications of the environments used to generate runtime data.}
    \footnotesize
    \centering
    \begin{tabular}{|c|l|l|}
        \hline
        Label & {Clock speed} & {OS (Compiler)} \\ \hline
        Win7 & i7-6800K (3.40 GHz) & Windows 7 (msvc 14.28) \\ 
        Win10 & i7-7700HQ (2.80 GHz) & Windows 10 (msvc 14.27) \\
        Arch & i5-6500 (3.20 GHz) & Arch Linux (gcc 10.2) \\
        \hline
    \end{tabular}
    \label{table:hardware}
\end{table}

\section{Requirements and Limitations}
\subsection{Hardware and Software Environment}

Table~\ref{table:compilers} lists compilers, operating systems and target architectures that were used in testing \SymPhas{}.
The minimum \CC{} standard is \CC{}17.
The minimum gcc version required to build \SymPhas{} is gcc7.5. This version notably introduces constructor template deduction guides, a necessary part of the compile time expression algebra.
The latest Microsoft Visual \CC{} Compiler (abbreviated as MSV\CC{} or MSVC) version is highly recommended. As a result of the heavy usage of meta-programming, earlier versions of MSV\CC{} are not guaranteed to successfully build \SymPhas{}.

When compiling \SymPhas{}, four of the modules are required for the minimum build and there is only one required external dependency, FFTW. These are listed in Table~\ref{table:dependencies} alongside the minimum tested versions.

\begin{table}
    \caption{The environments used for testing and compiling \SymPhas{}. The target architecture is specified in the third column. The minimum gcc version required to build \SymPhas{} is gcc7.5. The latest MSV\CC{} version is highly recommended. }
    \centering
\begin{tabular}{|l|l|l|l|}
     \hline
     Compiler & Operating System & Target Arch. & Compiles?  \\
     \hline
     MSV\CC{} 14.28 & Windows 7 Professional (64-bit) & x64 & Yes \\
     MSV\CC{} 14.28 & Windows 10 Home (64-bit) & x64 & Yes \\
     clang 11.0.1 & Arch Linux (64-bit) & x86-64 & Yes \\
     g++ 10.2 & Arch Linux (64-bit) & x86-64 & Yes \\
     g++ 7.5 & Arch Linux (64-bit) & x86-64 & Yes \\
     g++ 5.5 & Arch Linux (64-bit) & x86-64 & No \\
     \hline
\end{tabular}
    \label{table:compilers}
\end{table}

\begin{table}
    \caption{List of the dependencies of each module. Modules that are required in the base build of \SymPhas{} are emphasized using bold print. Optional modules are always an optional dependency. The version of the external dependency with which \SymPhas{} has been tested is indicated in parentheses. The dependency tbb enables parallelism when using the execution header.}
    \centering
    \begin{threeparttable}
        \begin{tabular}{|l|l|l|}
            \hline
            Module & Internal Dependency & External Dependency  \\ 
            \hline
            \textbf{\modulename{lib}} & None & FFTW (3.3.7) \cite{Frigo_2005}, tbb* \\
            \textbf{\modulename{datatypes}} & \modulename{lib} & None  \\ 
            \textbf{\modulename{sym}} & \modulename{datatypes} & None \\
            \textbf{\modulename{sol}} & \modulename{sym}, \modulename{io} & VTK \cite{vtk} (9.0)**  \\
            \modulename{io} & \modulename{datatypes} & libxdrfile (2.1.2)** \\
            \modulename{conf} & \modulename{sol}, \modulename{io} & None \\
            \hline
        \end{tabular}
        \begin{tablenotes}
            \item
            \small
            *Library is only required for compiling in Linux.\\
            **Library is optional.
        \end{tablenotes}
    \end{threeparttable}
    \label{table:dependencies}
\end{table}

\subsection{Limitations}

When the equations of motion and equations for virtual variables are written, a corresponding expression tree will be constructed by the symbolic algebra functionality. However, virtual variables used in the equations of motion will be substituted directly as data variables rather than as expression trees, meaning that the expression tree associated with the virtual variable will not be used to construct the expression tree for the equation of motion.
One implication of this design is that computing the derivative of a virtual variable that is defined in terms of a derivative will apply a stencil twice, resulting in a poor approximation. 
This also means that when using the spectral solver with equations of motion involving virtual variables, the spectral operators may be malformed if a virtual variable is written using a term necessary to correctly construct the operator. Refer to Section~\ref{sec:spectral} which explains the procedure of constructing the operators.

The Euler solver will assume that it can approximate derivatives of any order, but is only able to compute derivatives for which the stencils are implemented. See Section~\ref{methods:objects:stencils} for the exhaustive list of stencils.

Only scalar values can be provided to the model arguments for initializing the values of \lstinline{c1}, \lstinline{c2}, $\ldots$, and cannot be other types like matrices or complex types. When the model equations should use other types, then an appropriate number of scalar arguments should be provided in order to construct the term in the model preamble.

\section{Conclusions}

With \SymPhas{}, we have developed a high-performance, highly flexible API that allows a user to simulate phase-field problems in a straightforward way. This applies to any phase-field problem which may be formulated field-theoretically, including reaction-diffusion systems. Simulated models are written using the equations of motion in an unconstrained form specified through simple grammar
that can interpret mathematical constructs. Here, \SymPhas{} was tested in both 2- and 3-dimensions against the well-known Cahn--Hilliard~\cite{Cahn1958} and Allen--Cahn~\cite{Allen_1972} models, a model of two coupled equations of motion for eutectic systems~\cite{Elder1994}, and a phase-field crystal model~\cite{Elder2002}. The results demonstrate that \SymPhas{} produces correct solutions of phase-field problems.
Overall, \SymPhas{} successfully applies a modular design
and supports the user by providing individual modules for each functional component alongside a highly detailed documentation.

With the growing interest in phase-field methods outside traditional materials and microstructure modeling, subjects such as wave propagation in cardiac activity and properties of biological tissues are other potential application fields  \cite{Courtemanche_1996, Raina_2015, Gueltekin2016}. \SymPhas{}  offers
very short definition-to-results workflow 
facilitating rapid implementation of new models.
In addition to being a tool for direct simulations, \SymPhas{} can generate large volumes of training data for new machine learning simulations of phase-field models. For instance, this type of approach has been applied in recent works that focused on formulating a free energy description from the evolving microstructure \cite{Teichert_2019, Zhang_2020} and in developing machine-guided microstructure evolution models for spinodal decomposition \cite{OcaZapiain2021}.

\SymPhas{} will continue to be developed and have features added over time, which will include items such as upgrades to performance, additional symbolic algebra functionality, stochastic options and more solvers. The software is available at \texttt{https://github.com/SoftSimu/SymPhas}.

\medskip

\textbf{Supporting Information} \par
Supporting Information is available 
from the authors.

\medskip

\textbf{Conflict of Interest} \par The authors declare no conflict of interest.

\medskip
\textbf{Acknowledgments} \par 
M.K. thanks  the  Natural  Sciences  and  Engineering  Research  Council  of  Canada (NSERC) and Canada Research Chairs Program for financial support. 
S.A.S. thanks NSERC for financial support through the Undergraduate Student Research Award (USRA), the Canada Graduate Scholarships - Master's (CGSM) programs and Mitacs for support through the Mitacs Globalink Research Award. 
SharcNet  and  Compute  Canada  are acknowledged for  computational resources.

\medskip

\printbibliography

\end{document}